%% file: main.tex
\documentclass[journal]{IEEEtran}
\usepackage{cite}
\usepackage{amsmath,amssymb,amsfonts}
\usepackage{algorithmic}
\usepackage{graphicx}
\usepackage{textcomp}
\usepackage{xcolor}
\usepackage[normalem]{ulem}
\usepackage[nolist]{acronym}
\usepackage{verbatim}
\usepackage{subfigure}
\def\BibTeX{{\rm B\kern-.05em{\sc i\kern-.025em b}\kern-.08em
    T\kern-.1667em\lower.7ex\hbox{E}\kern-.125emX}}

\input{Definitions}
\input{Acronyms}

\begin{document}

\title{Performance Analysis of IEEE 802.11p Preamble Insertion in C-V2X Sidelink Signals\\for Co-Channel Coexistence}

\author{Alessandro Bazzi, 
Stefania Bartoletti, 
Alberto Zanella, 
and Vincent Martinez
\thanks{A. Bazzi is with Universit\`{a} di Bologna, Viale Risorgimento 2,
	40136 Bologna, Italy
	(e-mail: {\tt alessandro.bazzi@unibo.it}).}
\thanks{S. Bartoletti and A. Zanella are with CNR, Italy (e-mail: {\tt\{stefania.bartoletti,alberto.zanella\}@ieiit.cnr.it}).}
\thanks{A. Bazzi, S. Bartoletti, and A. Zanella are also with CNIT WILAB, Italy.}
\thanks{Vincent Martinez is with NXP, Toulouse, France (e-mail: {\tt vincent.martinez@nxp.com}).}
}

\maketitle

\begin{abstract}
Spectrum scarcity is one of the main challenges of future wireless technologies. When looking at \ac{V2X}, this is amplified as spectrum sharing could impact road safety and traffic efficiency. It is therefore of particular importance to study solutions that allow the coexistence, in the same geographical area and in the same channels, of what are today the main V2X access technologies, namely IEEE~802.11p and sidelink LTE-V2X Mode~4. 
In this work, in addition to investigating the impact of the reciprocal interference, which we demonstrate to have a strong impact especially on the first and in congested channel conditions, a mitigation solution is extensively studied, which is based on the insertion of the IEEE~802.11p preamble at the beginning of the LTE-V2X sidelink transmission. The proposal, which is also under discussion within the standardization bodies, requires no modifications to the IEEE~802.11p protocol stack and minor changes to LTE-V2X sidelink. This solution is directly applicable to upcoming IEEE 802.11bd and extendable to NR-V2X sidelink. 
The paper shows, through analysis and simulations in free-flow and dense scenarios, that the proposal allows for a mitigation of collisions caused by co-channel coexistence under low to high-load channel conditions and that the improvement is also granted in congested cases when combined with additional countermeasures. Regarding the latter aspect, in particular, different approaches are compared, demonstrating that acting on the congestion control mechanisms is a simple but effective solution.
\end{abstract}


\acresetall

\section{Introduction}

With the aim of improving road safety and traffic efficiency, spectrum has been reserved in several countries for
\ac{V2X} communications between
\acp{OBU} and \acp{RSU}. Since the beginning of 2021, the first cars equipped as standard with IEEE 802.11p, have
reached the consumer market.\footnote{In particular, the first models are sold in Europe since the beginning of 2021 with IEEE 802.11p as standard equipment and at that time over 6000~km of roads were already covered by commercially distributed RSUs (source: C-Roads at C2C-CC Forum, November 2020).} It is however clear that the radical change enabled by short-range connectivity will be visible only when the market penetration  increases significantly. 

Among the reasons for the delayed roll-out, although the solutions are available and tested on a large scale, there is the technical debate involving the two families of standards that have been defined for this purpose, namely the one based on IEEE~802.11 and the other based on sidelink technologies designed by the \ac{3GPP} for V2X in LTE and 5G. Given the failed attempts to impose one of the two technologies,\footnote{In the US, a proposed rule was issued in January 2017, which resulted in a stalemate. In Europe, the European Commission proposed a Delegated Act in March 2019, which was rejected by the European Parliament in July 2019.}, one of the main open issues is the investigation of their coexistence in the same channels, namely spectrum sharing, and solutions to mitigate the mutual interference.

The co-channel coexistence of different technologies is in general a well known and studied issue. For example, all technologies that use the \ac{ISM} bands must apply some mechanisms to limit the interference produced to possible coexisting technologies. Just to name a few, ZigBee and WiFi adopt a \ac{CSMA/CA} mechanism, Bluetooth implements \ac{FHSS}, and LoRa implements \ac{CSS}. Furthermore, proposals have been presented in the literature to further mitigate mutual interference, 
for example considering Wi-Fi and LTE in the unlicensed ISM bands \cite{6957143,8806680}.
However, co-channel coexistence represents a new topic when it comes to the ITS band, which is today associated with IEEE 802.11p and LTE-V2X sidelink (simply LTE-V2X below, for the sake of conciseness). As far as we know, this topic is  in fact taken into consideration only in some early works \cite{BazZanSarMar:C20, roux2020performance,9644653} and in the study published in ETSI TR 103 766 \cite{ETSI_TR_103_766}, where some solutions are proposed and preliminary investigated to reduce the impact of inter-technology interference.

In this paper, we propose and investigate a solution to mitigate collisions due to mutual interference between IEEE~802.11p and LTE-V2X communications, which consists in the insertion of a fixed preamble at the beginning of the LTE-V2X signal that improves the sensing capability of IEEE~802.11p. Because  this preamble sequence is fixed and predefined, LTE-V2X stations can use a recorded sequence and do not need to implement the IEEE~802.11p standard. In addition, this solution is compatible with IEEE 802.11bd and can be extended to 5G-V2X sidelink.

Although the insertion of the preamble is
part of one of the methods discussed in the ETSI technical report  \cite{ETSI_TR_103_766}, where it is used together with a time domain sharing mechanism that distinguishes between LTE-V2X and ITS-G5 slots, this mechanism has never been proposed as a stand-alone coexistence mitigation solution and a formal analysis of its implications has not been provided in the previous works.

The paper is organized as follows. Having recalled the main aspects of both technologies and discussed the implications of their co-channel coexistence in Section~\ref{sec:techs}, we describe in detail the concept of preamble insertion in Section~\ref{sec:preamble}. We then investigate its performance through Sections~\ref{sec:resSimplified} to~\ref{Sec:congested}. In particular:
\begin{itemize}
    \item We first focus on a free-flow scenario in Section~\ref{sec:resSimplified} and introduce  a  mathematical  model  for  the  analysis  of the IEEE 802.11p preamble insertion, showing the significant reduction of inter-technology collisions;
    \item Then, we study denser scenarios in Section~\ref{sec:resSystem} through simulations obtained with an open source software, which confirm the validity of the approach for medium and high traffic scenarios;
    \item Finally, we address the congested scenarios in Section~\ref{Sec:congested} and compare three possible additions to dealing with them, demonstrating that acting on the congestion control procedures of the LTE-V2X sidelink is more effective than the other solutions.
\end{itemize} 
Our conclusions are finally drawn in Section~\ref{sec:conclusion}.

\section{Technologies and Coexistence Issues in Brief}\label{sec:techs}

This section briefly describes the two technologies and discusses the issues that arise when they are adopted in the same channel in the same geographical area.

\subsection{IEEE 802.11p and related standards} 

IEEE 802.11p, completed in 2010 and now part of IEEE 802.11-2020 \cite{IEEE80211_2020}, is based on \ac{OFDM} at the \ac{PHY} layer and \ac{CSMA/CA} at the \ac{MAC} layer. In the US, it is used for the lower layers of the protocol stack referred to as \ac{WAVE}, which includes the IEEE 1609 standards and is supplemented at the upper layers by SAE documents. In Europe, it is used for the ITS-G5 access layer, which has been defined by ETSI along with a number of standards that cover all the layers of the protocol stack. Currently, an evolution is being defined as IEEE 802.11bd, whose publication is expected by end of 2022.

\begin{figure} [t]
	\centering
	\subfigure[IEEE 802.11p.]{
		\includegraphics[width=0.92\columnwidth]{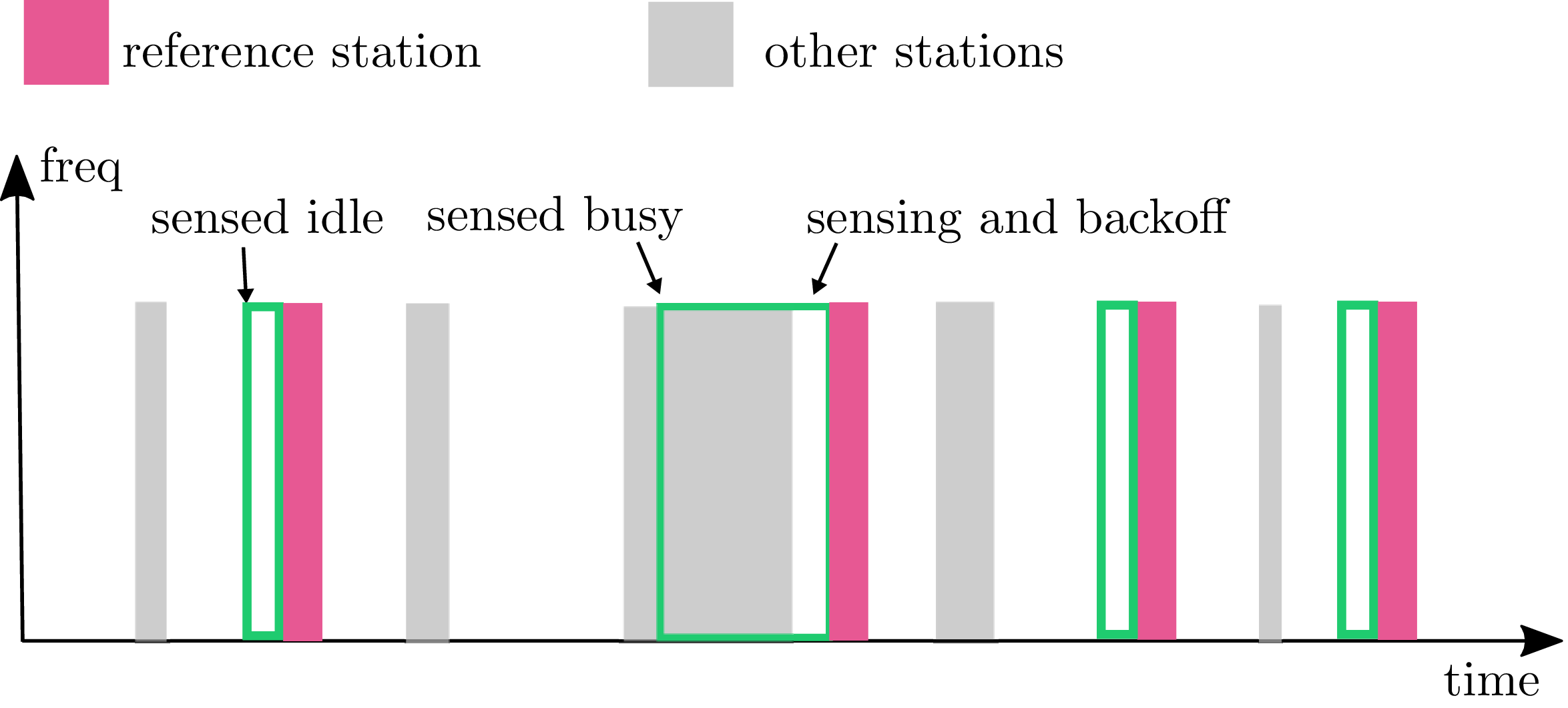}\label{fig:ITS-G5}}
	\subfigure[LTE-V2X sidelink Mode 4.]{
		\includegraphics[width=0.92\columnwidth]{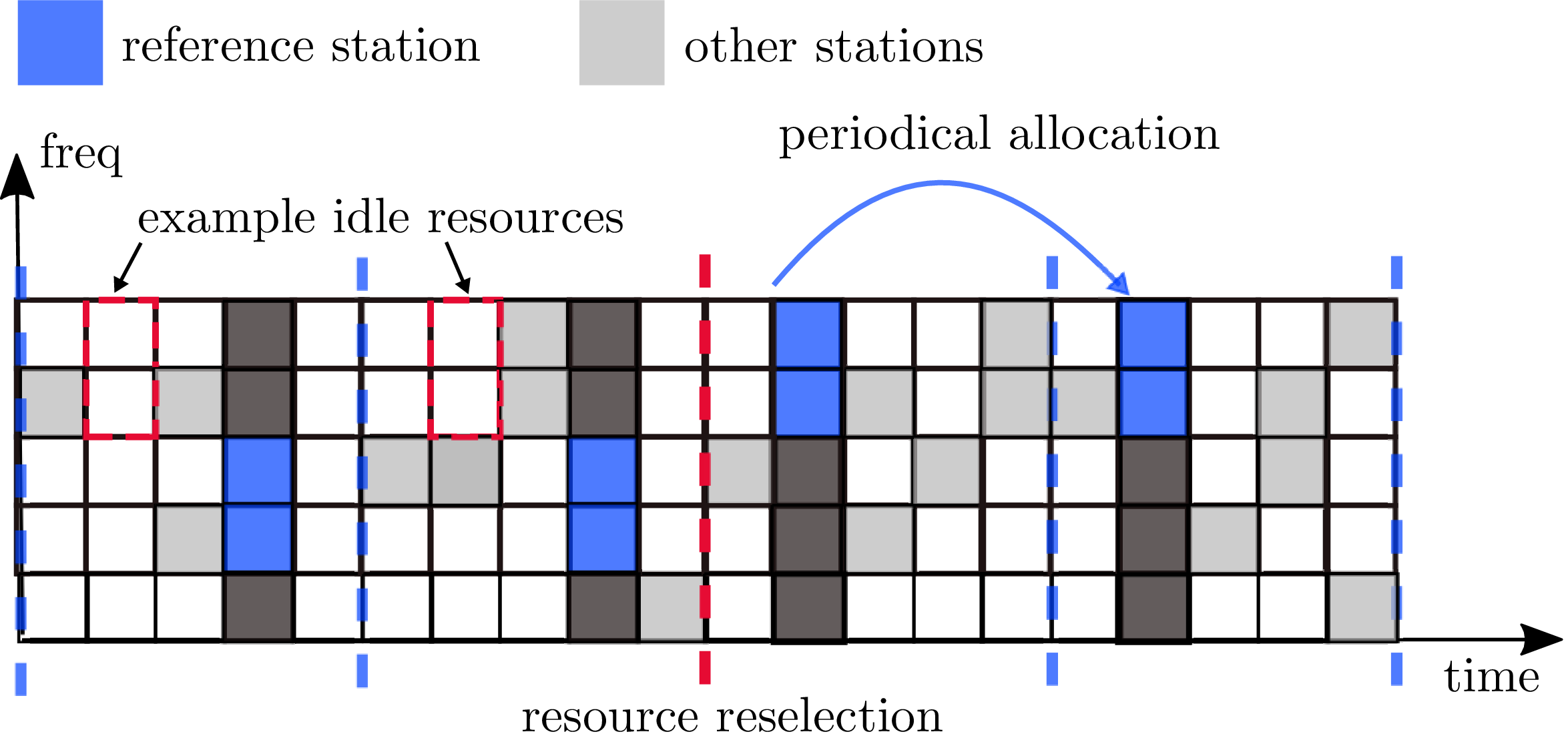}\label{fig:LTEV2X}}
	\caption{Illustration of the medium access mechanisms.}
	\label{fig:MAC}\vskip -0.3cm
\end{figure}	

IEEE 802.11p, with its \ac{CSMA/CA} \ac{MAC}, is a fully distributed asynchronous ad hoc technology. Each time a station has new data to transmit, it senses the medium for a certain duration and starts the transmission if the channel is idle, otherwise it postpones it. When the transmission starts, all the subcarriers are used for a time that depends on the size of the packet and the \ac{MCS} adopted. In \ac{V2X}, for the moment frames are sent in broadcast mode and therefore there is no acknowledgment returned by the receiver(s).

The access mechanism of IEEE 802.11p and the use of the channel are illustrated in Fig.~\ref{fig:ITS-G5}. More details about IEEE~802.11p can be found for example in \cite{SepGozCol:J17,CamMolVinZha:J12}.

\begin{figure*} [th!]
	\centering
	\subfigure[Legacy.]{
		\includegraphics[width=0.4\linewidth,draft=false]{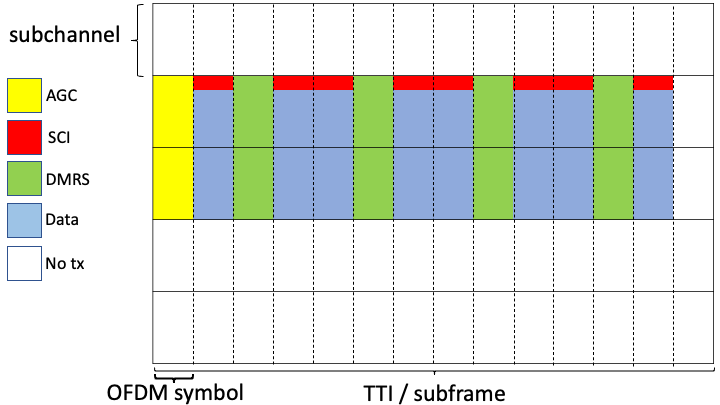}\label{fig:symbolsa}}~~~~~~~~
	\subfigure[With IEEE~802.11p preamble insertion.]{
		\includegraphics[width=0.4\linewidth,draft=false]{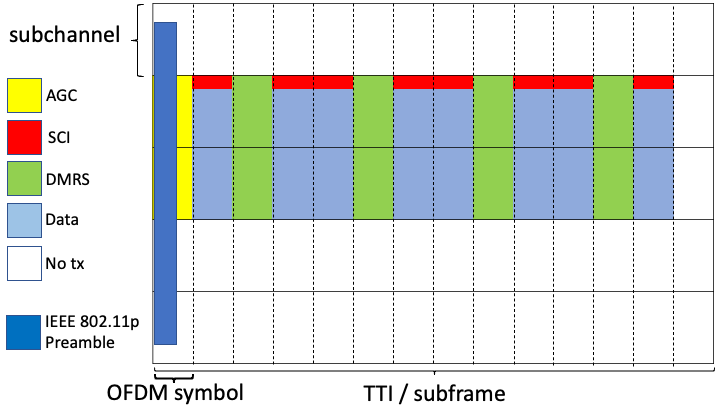}\label{fig:symbolsb}}
	\caption{Comparison between legacy and preamble insertion. Example of an LTE-V2X transmission using two out of the five subchannels, assuming adjacent allocation of the sidelink control information. Each figure represents a signal using two out of the five subchannels for the 14 OFDM symbols that form one time transmission interval. AGC for automatic gain control, SCI for sidelink control information, DMRS for demodulation reference signals.}
	\label{fig:symbols}\vskip -0.3cm
\end{figure*}

\subsection{C-V2X Sidelink and LTE-V2X Mode 4} 

Since Release 14, frozen in 2016, 3GPP has introduced LTE-V2X, referred to as part of the so-called \ac{C-V2X}. \ac{C-V2X} is in fact an umbrella that covers both LTE and 5G, and includes both downlink/uplink (Uu interface) and sidelink (PC5 interface) communications. Focusing on the sidelink, resource allocation can be performed from the network, called Mode~3 in LTE and Mode~1 in 5G, or autonomously from stations, called Mode~4 in LTE and Mode 2 in 5G. In this article, LTE-V2X sidelink Mode~4 is considered, as it is the solution currently under consideration for the ITS band.

LTE-V2X is based at the \ac{PHY} and \ac{MAC} layers on \ac{SC-FDMA}, with a resource granularity equal to the \ac{TTI} in the time domain and the subchannel in the frequency domain. One TTI, also called subframe, is 1~ms long, while one subchannel occupies a predefined number of \acp{PRB} \cite{BazZanMas:J19}, with a PRB corresponding to 12 subcarriers occupying 180~kHz. Each transmission therefore lasts 1~ms and occupies a portion of the bandwidth that depends on the payload size and the \ac{MCS} adopted. In LTE-V2X, each message can be transmitted twice thanks to the so-called \ac{HARQ}, which is a blind retransmission mechanism. 

The allocation process performed at the MAC layer with Mode~4 is designed by assuming periodic messages and is based on a mechanism called \ac{SB-SPS}: a station with a packet to be transmitted, estimates the resource occupancy in the last time window of 1~s, and based on these measurements as well as information from the \ac{SCI} associated with each packet, deduces what the occupation will be in the future. Once the resource is allocated, no additional channel measurement is made before the packet is transmitted. The same resource is then used periodically for a certain duration before repeating the procedure. 

The details of Mode 4, which is exemplified in Fig.~\ref{fig:LTEV2X} and has been extensively  studied in recent years, can be found for example 
in \cite{BazCecZanMas:J18,MolGozSep:C18,TogSaiMahMugFalRaoDas:C18}. 

\subsection{Co-channel coexistence} 

When looking at co-channel coexistence, we can note that
IEEE~802.11p, through the use of \ac{CSMA/CA}, also known as \textit{listen-before-talk}, is inherently designed to limit its interference to other technologies. Differently, LTE-V2X comes from technologies that normally do not need to cope with co-channel coexistence and are not designed to take  this aspect into account. As already noted, an LTE-V2X station adopting Mode~4, starts a transmission using the previously allocated resources, without any additional verification.
The problem is further exacerbated in very loaded channel conditions, with the risk of strong unfairness due to the fact that LTE-V2X tends to use most of the resources, while IEEE 802.11p tends to postpone and eventually reduce its access.

Another possible issue, worthy of note even if not analysed in this work, is caused by the last OFDM symbol of the LTE-V2X subframe, which is left empty to allow the switching from the transmission phase to that of reception, therefore leading to about 71.4~$\mu$s of idle channel between two consequent LTE-V2X transmissions. In some cases,\footnote{For example with the parameters defined in Europe for high-priority \acp{DENM} in ITS-G5.} this duration could be longer than the access time of IEEE 802.11p and the sensing mechanism of its CSMA/CA would not be able to detect in time the use by the LTE-V2X stations of the subsequent TTI.

\section{IEEE 802.11p Preamble Insertion}\label{sec:preamble}

An IEEE~802.11p transmission starts with a preamble of 40~$\mu$s, which includes a short and a long training sequence, each lasting for 16~$\mu$s, and the Signal Field OFDM symbol, lasting 8~$\mu$s. This symbol adopts BPSK and 1/2 coding rate, carrying 24 useful bits on 48 data subcarriers. From these 24 bits, the duration of the remaining signal can be derived. Note that two signals of the same duration will begin with exactly the same preamble.

In legacy LTE-V2X, the signal, which occupies one TTI of 1~ms, is divided into 14 OFDM symbols of approximately 71.4~$\mu$s. As illiustrated in Fig.~\ref{fig:symbolsa}, the first symbol is used for \ac{AGC} (it carries a copy of the second symbol), 4 are used for \acp{DMRS}, 8 for data, and the last one is left empty to allow switching from transmission to reception mode. In Fig.~\ref{fig:symbolsa}, the channel is organized into five subchannels in accordance with \cite{ETSI_TS_103_613} and, as an example, the signal occupies two of them. 

The idea of the IEEE 802.11p preamble insertion is to replace, as shown in Fig.~\ref{fig:symbolsb}, the first 40~$\mu$s of the LTE-V2X signal with an IEEE~802.11p preamble indicating the occupation of the channel for 1~ms.\footnote{A different proposal is to place the header in the empty symbol prior to the subframe.} As mentioned, the first part of the LTE-V2X signal is anyway carrying redundancy and the added preamble is always the same, thus stored IQ samples could be used to generate the signal without additional complexity to the LTE-V2X transmitter. The AGC is still performed, although using in part a different signal, and the data-rate of LTE-V2X is not reduced. Also relevant to note is that the LTE-V2X station does not need to implement the IEEE~802.11p standard. Since the preamble signal is always the same for any station at any point of time, multiple LTE-V2X signals transmitted simultaneously will have the same impact on the decoding of the preamble of multiple paths.

Two advantages are obtained with the considered approach. First, the LTE-V2X occupancy of 1 ms is indicated as part of the \textit{signal field} of the preamble and therefore the gap at the end of the TTI does not cause CSMA/CA to sense the channel as idle. Second, more importantly, it significantly reduces the received power at which IEEE~802.11p assumes the channel as busy; in fact, from the -65~dBm of the \ac{CCA} threshold, which is used when an undecodable signal is received, the minimum power is reduced to a level that depends on the implemented receiver and can be reasonably assumed around -100~dBm.\footnote{The value of -100~dBm, which is in agreement with what obtained in off-the shelf devices,  corresponds to a \ac{SNR} of -2~dB with a noise figure of 6~dB. A slightly higher value is used here for performance evaluation, as motivated in Section~\ref{Subsec:freeflowresults}. Example reference is the Cohda MK5, fccid.io/2AEGPMK5RSU/Users-Manual/User-Manual-2618067.pdf} Please note that the simple reduction of the \ac{CCA} threshold to a lower level would not lead to the same result; the threshold has been optimized to avoid false detection of unexpected signals, such as spurious emissions from adjacent channels.

A difference from the current specification is that the addition of the preamble, as shown in Fig.~\ref{fig:symbolsb}, causes some power to be transmitted over all subchannels at the beginning of the TTI, even when part of the bandwidth is used for the rest of the TTI. This, in principle, might alter the \ac{SB-SPS} process of LTE-V2X. However, the signal is only transmitted for 40~$\mu$s over the 1~ms TTI and therefore the impact is negligible, as demonstrated in Section~\ref{subsec:techalone}.

A further advantage of the proposed solution is its applicability to IEEE 802.11bd and sidelink 5G-V2X. In IEEE 802.11bd, the same preamble is already part of the specifications by design. In the case of 5G-V2X, nothing is expected to change from LTE-V2X if the same numerology is used, i.e., with the same subcarrier spacing of 15~kHz and the same TTI of 1~ms. The methodology is also applicable when the subcarrier spacing is increased to 30~kHz and the TTI reduced to 0.5~ms: in that case, each OFDM symbol lasts about 35$\;\mu$s, but part of the gap from the previous symbol could be used to accommodate the preamble of 40\;$\mu$s. Only if a higher subcarrier spacing is used, which is not however planned for channels of 10 MHz in the ITS band, more than one symbol would be needed, which could cause a reduction in the useful data-rate.

\section{Impact in the free-flow scenario}\label{sec:resSimplified}

In this section, the impact of inserting the preamble is studied in a low-traffic scenario. For this purpose, a model is developed that allows to analyze the impact of the proposed solution on the collisions between transmissions from IEEE 802.11p and LTE-V2X coexisting in the same channel.

\subsection{Scenario and assumptions}\label{subsec:modelassumptions}

Given the low density, the free-flow scenario is reproduced by focusing on a single IEEE 802.11p transmission interfered by a variable number of LTE-V2X transmissions. The impact of the preamble insertion is here assessed in terms of \ac{PRP} of the IEEE 802.11p link, which is the technology most affected by the co-channel interference \cite{BazZanSarMar:C20,ETSI_TR_103_766}. Additional performance metrics are evaluated in Section~\ref{sec:resSystem} through simulations in denser scenarios. 

The instant when the IEEE 802.11p packet reaches the access layer transmission buffer is chosen randomly and not aligned with the LTE subframe structure; the time required for the trasnmission of a packet plus the preceding \ac{AIFS} is assumed to last for less than 1~ms.\footnote{1~ms corresponds to approximately 700 bytes adopting the default \ac{MCS}~2.} 
The performance is calculated in terms of \ac{PRP} by varying the source-destination distance of the IEEE 802.11p link and the average number of LTE-V2X transmissions per meter per second. 

\begin{figure} [t]
	\centering
	\subfigure[Example with both 2-D and 1-D representation.]{\includegraphics[trim= {0 200 0 0}, clip,width=0.98\linewidth,draft=false]
{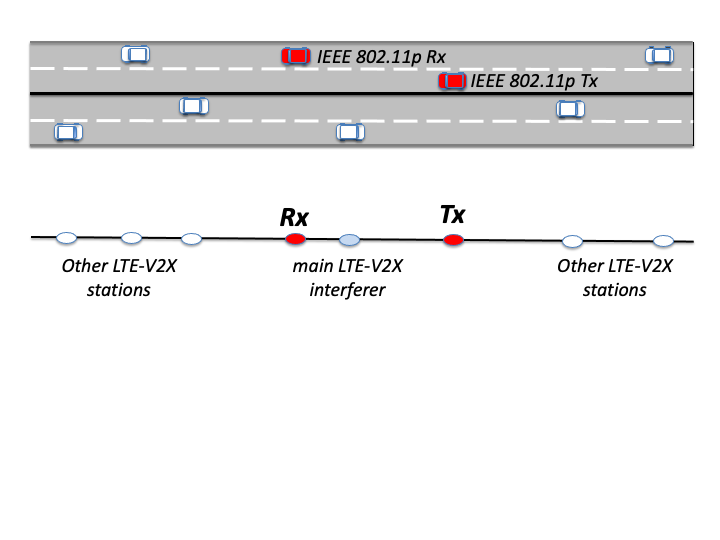} \label{Fig:AnalysisScenarioa}}
	\subfigure[Corresponding parameters.]{\includegraphics[trim= {0 20 0 250}, clip,width=0.98\linewidth,draft=false]
{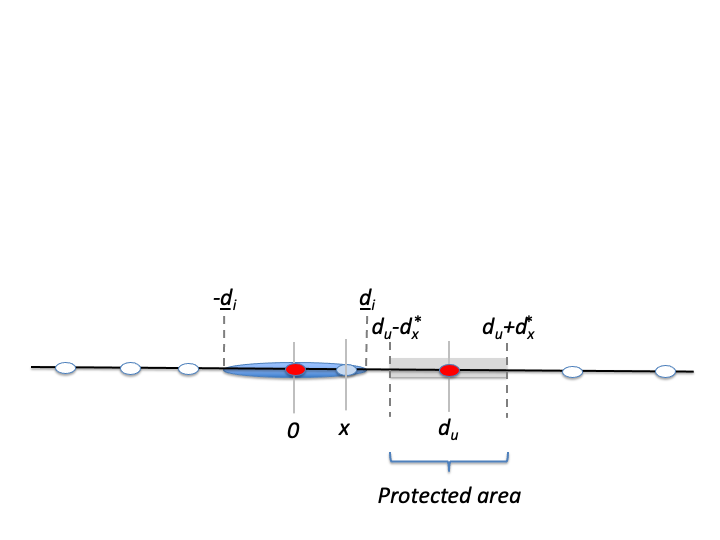} \label{Fig:AnalysisScenariob}}
	\caption{Free-flow scenario and parameters used in the model.}
	\label{fig:AnalysisScenario} \vskip -0.3cm
\end{figure}

The model is based on the following approximations: 
\begin{itemize}
    \item As represented in Fig.~\ref{Fig:AnalysisScenarioa}, the highway scenario is approximated as a straight line, so the LTE-V2X nodes correspond to a 1-D \ac{PPP} distribution (this approximation is used in several similar works, such as \cite{tong16,ZhaCheYanEtAl:J12,BazZanCecMas:J19}); 
    \item Since the scenario is low density, only the strongest source of LTE-V2X interference is considered in each subframe (also this approximation is adopted by several articles, such as \cite{ParKimHon:J18}); 
    \item The correct reception is modeled through a threshold model, which means that the packet is correct when the \ac{SINR} is above the threshold and is incorrect when it is below; fading effects are included in the threshold setting. The analysis is validated by simulations in which the channel is modeled in more details and includes log-normal large-scale fading (shadowing), and \ac{PER} vs. \ac{SINR} curves that account for small-scale fading.
\end{itemize}

\subsection{Model definition}

Considering a 1-D scenario as represented in Fig.~\ref{Fig:AnalysisScenariob}, the IEEE~802.11p receiver is assumed at position 0. Without lack of generality, the IEEE~802.11p transmitter is located on the right of the receiver, at position $\du$, i.e. the distance between the transmitter and the receiver is $\du$. There are on average $\averageLTEtx$ LTE-V2X transmissions per meter per second.

Since retransmission is not allowed in IEEE 802.11p, assuming interference from the generic LTE-V2X signal as white and Gaussian, the \ac{PRP} can be written as
\begin{equation}\label{eq:PRRI}
\PRR = 1-\fPER{\SINRu}
\end{equation}
where $\fPER{\SINRu}$ is the PER derived from the average \ac{SINR}, calculated as $\SINRu=\Pru/\left(\Pn+\PI\right)$, 
$\Pru=\Ptxp\Gt\Gr/\Loss(\du)$ is the average received power, $\Ptxp$ is the transmission power of IEEE~802.11p stations, $\Gt$ and $\Gr$ are the transmitter and receiver antenna gains, respectively (assumed the same for all transmitters and receivers for simplicity), $\Pn$ is the average power of noise, and $\PI=\Ptxlte\Gt\Gr/\Loss(d_\text{i})$ is the average power of interference from an LTE-V2X node that transmits with power $\Ptxlte$ at a distance $d_\text{i}$ from the receiver. For the sake of conciseness, in \eqref{eq:PRRI} $\du$ and $d_\text{i}$ are left implicit in $\SINRu$. 

The packet whose outcome is evaluated is generated in a generic instant, which results within the LTE subframe that we call the \textit{current \ac{TTI}}. The following LTE subframe is hereafter called \textit{subsequent TTI}. 
If the channel is sensed idle by the IEEE 802.11p station and the transmission begins, depending on the generation instant the transmission either is performed entirely within the current TTI or it ends during the subsequent TTI. 
Denoting as $\tPack$ the duration of the IEEE 802.11p transmission, including the \ac{AIFS}, and as $\tTTI$ the duration of one TTI, the transmission is entirely contained in the current TTI with probability $\Pwithin=\left(\tTTI-\tPack\right)/\tTTI$, and partially occupies the subsequent TTI with probability $\Pcross=1-\Pwithin$.

Since the position of the interfering LTE-V2X node and the TTI it uses are independent, the distribution of nodes using the current TTI is a \hbox{1-D} PPP process with density $\averageLTEtxN = \averageLTEtx/\NttiSec$, where $\NttiSec = 1 /\tTTI$ is the number of TTIs in one second. Similarly, the distribution of nodes using the subsequent TTI is also a \hbox{1-D} PPP process with density $\averageLTEtxN$, independent from the previous one.

For the properties of the PPP distributions, the strongest LTE-V2X interferer in the current TTI is in the position $\dIfirst$ (positive or negative), therefore at distance $|\dIfirst|$ from the destination, with probability $\Ppos(\dIfirst)=\averageLTEtxN e^{-2\averageLTEtxN |\dIfirst|}$. Similarly, the strongest LTE-V2X interferer in the subsequent TTI is in the position $\dIsecond$ with probability $\Ppos(\dIsecond)=\averageLTEtxN e^{-2\averageLTEtxN |\dIsecond|}$. 

We call the maximum distance at which the interfering signal can be sensed by the IEEE~802.11p transmitter as \textit{protected range} and denote it as $\dmaxx=\Loss^{-1}(\Prminx)$, where $\Prminx$ is the minimum received power to set the channel as busy. The area from $\du-\dmaxx$ to $\du+\dmaxx$ is called \textit{protected area} (see Fig.~\ref{Fig:AnalysisScenariob}). A transmission from an LTE-V2X node in the protected area is detected by the IEEE~802.11p transmitter during the carrier sense procedure; hence, in the case the main LTE-V2X interferer of the current TTI is within the protected area, the IEEE 802.11p station defers its transmission. 

Depending on the position of the main LTE-V2X interferer in the current TTI and the instant when the IEEE 802.11p transmission ends (same or subsequent TTI), three cases are possible: (i) the LTE-V2X interferer is within the protected area, in which case the transmission is deferred to the first TTI during which the main LTE-V2X interferer is outside the protected area; this event occurs with probability $\Psensed$ and causes a \ac{PRP} equal to $\PRRbusy$; (ii) the LTE-V2X interferer is outside the protected area and the IEEE 802.11p transmission ends within the current TTI; this event occurs with probability $\Pnotswithin$ and causes a \ac{PRP} equal to $\PRRcurrent$; and (iii) the LTE-V2X interferer is outside the protected area and the IEEE 802.11p transmission ends in the subsequent TTI; this event occurs with probability $\Pnotscross$ and causes a \ac{PRP} equal to $\PRRsubseq$. By the law of total probability, it is 
\begin{align}\label{eq:PRRcompleteStarting}
    \PRR =& \Psensed\PRRbusy + \Pnotswithin\PRRcurrent + \Pnotscross\PRRsubseq \;.
\end{align}

The probability that the channel is sensed busy by the IEEE 802.11p transmitter during the current TTI is 
\begin{align}\label{eq:Pbusy}
    \Psensed = & \int_{\du-\dmaxx}^{\du+\dmaxx} \Ppos(x) dx = \frac{1}{2} \left[ (1-e^{-2\averageLTEtxN(\dmaxx+\du)}) \right. \nonumber \\ 
    & \left. + \text{sign}(\dmaxx-\du) \cdot (1-e^{-2\averageLTEtxN|\dmaxx-\du|})\right]
\end{align}
where $\text{sign}(x)$ is the sign function, returning $+1$ if $x\geq0$ and $-1$ if $x<0$. The sign function and the absolute value in the second term of \eqref{eq:Pbusy} take into account the possibility that the protected area is partly in the negative axis (i.e., $\dmaxx>\du$), or not.
If the current TTI is sensed busy, the IEEE 802.11p station defers its transmission to the first TTI during which the main LTE-V2X interferer is outside the protected area. Given the independence of the distribution of LTE-V2X nodes in the TTIs, in this case the \ac{PRP} is equal to
\begin{align}\label{eq:PRRsubseqApproxCF}
    \PRRbusy=&
    \frac{1/2}{1-\Psensed} \left[ (e^{-2\averageLTEtxN\cdot\left(\max\{\dImin,(\dmaxx-\du)\}\right)}) \right.\nonumber \\ & \left. + (e^{-2\averageLTEtxN\cdot\left(\max\{\dImin,(\dmaxx+\du)\}\right)})\right]
\end{align}
where $\max\{x,y\}$ is a function that returns the maximum between $x$ and $y$, and $\dImin$ is the minimum distance corresponding to the maximum interference to receive the packet correctly; note that while $\dmaxx$ is independent on $\du$, $\dImin$ varies with $\du$. The derivation of \eqref{eq:PRRsubseqApproxCF} is detailed in Appendix A.

The same \ac{PRP} is obtained also if the reference transmitter is not able to sense the LTE signal in the current TTI and ends in the current TTI. 
Such an event, occurring with probability 
\begin{align}\label{eq:Pnotswithin}
    \Pnotswithin = \left( 1-\Psensed \right) \Pwithin
\end{align}
is therefore characterized by an PRP equal to 
\begin{align}\label{eq:prrcurrent}
    \PRRcurrent=\PRRbusy\;.
\end{align}

In case the reference transmitter is not able to sense the LTE signal in the current TTI and ends in the subsequent TTI, which occurs with probability 
\begin{align}\label{eq:pnotcross}
\Pnotscross=1-\Psensed-\Pnotswithin = \left(1-\Psensed\right) \left(1-\Pwithin\right)\;,
\end{align}
\ac{PRP} is a function of both the interference in current TTI and that in the subsequent TTI. Note that the ability of the reference transmitter to sense or not LTE transmissions in the subsequent TTI is irrelevant, since the IEEE~802.11p transmission has already been started when the LTE-V2X transmission begins. The exact expression is given in Appendix B and includes a triple integral. However, by approximating the interference as entirely caused by the LTE-V2X transmission that overlaps more with the reference transmission, we obtain 
\begin{align}\label{eq:PRRsubseqApprox}
    \PRRsubseq \simeq \frac{\PRRbusy}{2} + \frac{ \PRRunprotected}{2}
\end{align}
having defined with
\begin{align}\label{eq:prrunprotectedCF}
\PRRunprotected=e^{-2\averageLTEtxN\dmaxx}
\end{align}
the PRP in the presence of interference without the sensing procedure (thus, unprotected). The derivation of \eqref{eq:PRRsubseqApprox} and \eqref{eq:prrunprotectedCF} is detailed in Appendix B.

As a consequence of \eqref{eq:Pnotswithin}, \eqref{eq:prrcurrent}, \eqref{eq:pnotcross}, and \eqref{eq:PRRsubseqApprox}, the \ac{PRP} in \eqref{eq:PRRcompleteStarting} can be rewritten as
\small
\begin{align}\label{eq:PRRcomplete}
    \PRR \simeq& \Psensed\PRRbusy + \Pnotswithin\PRRbusy + \Pnotscross\left(\frac{\PRRbusy}{2} + \frac{ \PRRunprotected}{2} \right)\nonumber \\ =& \left(1-\frac{\Pnotscross}{2}\right)\PRRbusy + \left(\frac{\Pnotscross}{2}\right) \PRRunprotected\;.
\end{align}
\normalsize

By using the results of \eqref{eq:Pbusy}, \eqref{eq:PRRsubseqApproxCF}, and \eqref{eq:prrunprotectedCF}, \eqref{eq:PRRcomplete} gives a closed-form expression of the \ac{PRP} as a function of $\du$ and $\lambda$.

\input{TableSettings.tex}

\begin{figure} [t]
	\centering
\subfigure[Packet reception probability of the IEEE 802.11p link vs. the link distance, with 1000 LTE-V2X transmissions per kilometer per second.]{\includegraphics[width=0.95\linewidth,draft=false]{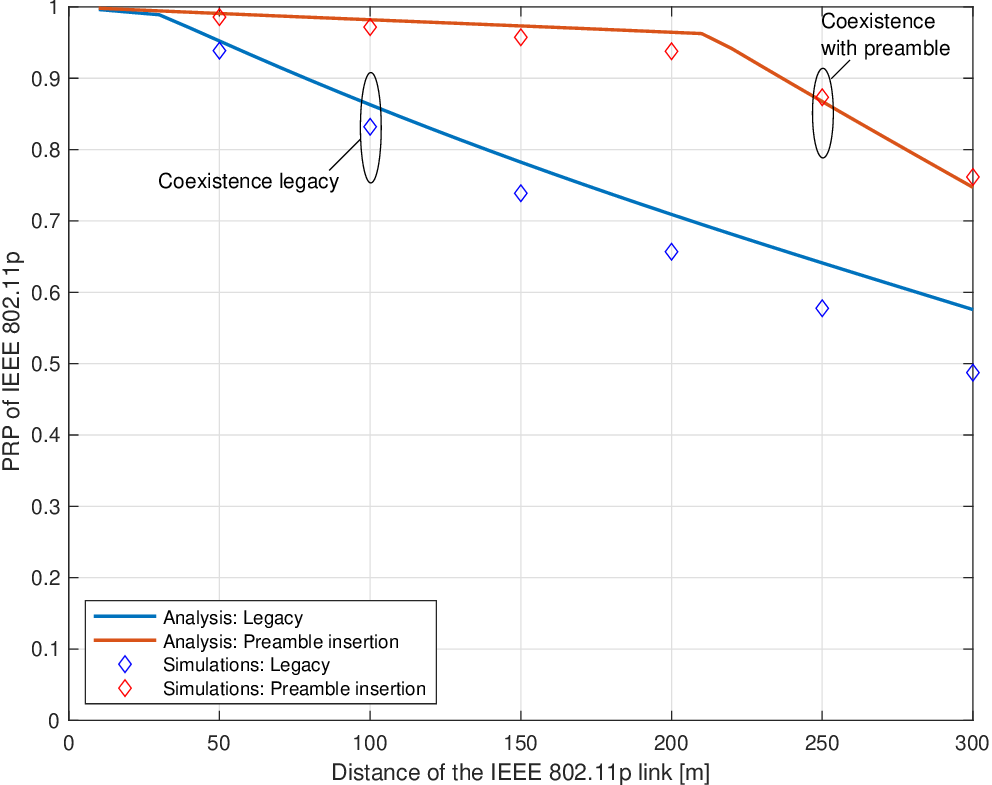}\label{fig:analysisPRRvsDistance}}
\subfigure[Packet reception probability of the IEEE 802.11p link at 200\;m vs. average LTE-V2X transmissions per kilometer per second.]{\includegraphics[width=0.95\linewidth,draft=false]{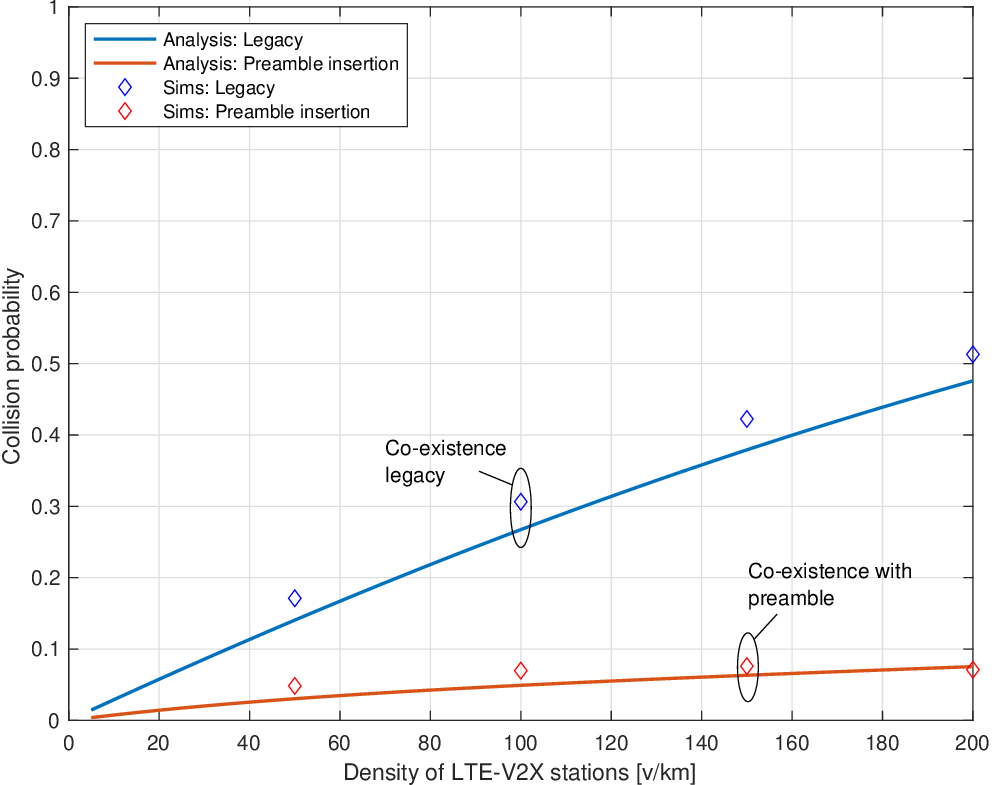}\label{fig:analysisPRRvsDensity}}
	\caption{Free-flow scenario. Impact of the preamble insertion on the packet reception probability of one IEEE 802.11p transmission interfered by coexisting LTE-V2X transmissions.
	}
	\label{fig:analysis}\vskip -0.3cm
\end{figure}

\subsection{Results in the free-flow scenario}\label{Subsec:freeflowresults}

Fig.~\ref{fig:analysis} shows the performance results obtained using the proposed model and assuming the settings listed in Table~\ref{Tab:Settings}. The $\SINRt$ adopted in the analysis corresponds to the \ac{SINR} of the \ac{PER} vs. \ac{SINR} curve used in the simulations with PER equal to 0.5. The WINNER+, scenario B1 model is used for the path-loss (as suggested by the 3GPP in \cite{3GPP_TR_36_885}), which is approximated in the analysis as $\Loss(\du)[dB] = \alpha + 10\cdot\beta\cdot\log_{10}(\du)$, where $\alpha=20.06$~dB and $\beta=4$. 

In the simulations used to validate the model and for the results discussed in the following sections, for each packet potentially received, the average \ac{SINR} is calculated taking into account the average interference from all other nodes, with the useful and interfering power calculated taking into account path-loss and log-normal correlated large-scale fading (shadowing). Once SINR is calculated, the correct reception of each packet is statistically drawn by the \ac{PER} vs. SINR curves shown in \cite{BazZanSarMar:C20}, which account for the impact of small-scale fading. More details can be found in \cite{TodBarCamMolBerBaz:21}.

To take into account that the preamble is more protected than the packet, it is assumed correctly decoded when the \ac{SINR} is above the value in the curves used for data corresponding to 0.9 PER; this means approximately -0.8~dB SINR and thus a minimum received power of about -98.8~dBm to successfully decode the preamble in the absence of interference. This value is in good agreement with common IEEE 802.11p receivers. 
The protected range corresponds to about 55~m in the legacy case, due to $\Prminx=-65$~dBm, and about 390~m with the addition of the preamble, due to $\Prminx=-98.8$~dBm.

In Fig.~\ref{fig:analysisPRRvsDistance}, the \ac{PRP} of the IEEE 802.11p link is shown by varying the transmitter-receiver distance with 1000 LTE-V2X transmissions per kilometer per second (which corresponds, for example, to 50 vehicles per kilometer that generate a message every 100\;ms and transmit it twice with blind retransmissions). The coexistence between the two standard (legacy) technologies is compared with that of LTE-V2X with the preamble insertion and standard IEEE 802.11p. Despite the approximations adopted, analysis and simulations show a very similar trend. Looking at Fig.~\ref{fig:analysisPRRvsDistance} and comparing the \ac{PRP} with legacy and with preamble insertion, the improvement due to the insertion of the preamble is clear. The PRP improvement granted by the preamble insertion begins to reduce at almost 200~m, which roughly corresponds to the distance by which $\dImin$ equals $\dmaxx-\du$. This value impacts directly on \eqref{eq:PRRsubseqApproxCF}: until $\dImin<\dmaxx-\du$, the interferer is located in the protected area, thus the channel is sensed as busy and the transmission is deferred; when $\dImin>\dmaxx-\du$, the channel is sensed as idle and the transmission begins immediately, possibly ending in a collision.

In Fig.~\ref{fig:analysisPRRvsDensity}, the \ac{PRP} at 200\;m is shown by varying the density of LTE-V2X transmissions. Analysis and simulations still show a similar trend. Comparing performance with legacy and preamble insertion, the success of the proposed approach to reducing the effect of LTE-V2X interference on the IEEE 802.11p transmission is again apparent. Implicitly, the results indicate a reduction in overlap between IEEE 802.11p and LTE-V2X signals, which also improves LTE-V2X reliability, as confirmed in the next section.

\section{Impact in denser scenarios}\label{sec:resSystem}

In this section, the study is expanded to more complex scenarios, through the use of the open-source simulator WiLabV2Xsim \cite{TodBarCamMolBerBaz:21}.\footnote{The simulator is available at https://github.com/V2Xgithub/WiLabV2Xsim. Modifications made for this study will be included in future releases of the simulator and in the meanwhile provided on request.} 
The main settings are detailed in Table\;\ref{Tab:Settings}. 

The following metrics are considered for both technologies:
\begin{itemize}
	\item \textit{Packet reception ratio (PRR)\acused{PRR}}, calculated as the average ratio between the number of vehicles at a certain distance from the transmitter that correctly decode a packet and the total number of vehicles at the same distance; PRR is calculated with a granularity of 10~m;
	\item \textit{Data age (DA)\acused{DA}}, corresponding to the time elapsed from the generation of a correctly received packet and the subsequent correctly decoded by the same receiver from the same transmitter; this metric includes the allocation delay and the correlation between errors; DA is evaluated for all transmissions within 400~m.
\end{itemize}

\subsection{Considerations with either technology alone}\label{subsec:techalone}

The proposed method has no impact if IEEE 802.11p is the only technology present. Differently, adding the preamble could, in principle, affect LTE-V2X behavior even when present alone. In fact, inserting the preamble implies that a certain power is distributed over the entire bandwidth at the beginning of an LTE-V2X signal, regardless of how many subchannels the rest of the signal uses (see Fig.~\ref{fig:symbols}). Therefore, at the beginning of an LTE transmission,  the nearby LTE-V2X stations detect a small increase in power even in the subchannels of the same TTI that are not used. In turn, the added power on the unused subchannels could in principle affect the \ac{SB-SPS} process of LTE-V2X.

\begin{figure} [t]
	\centering
	\includegraphics[width=0.95\linewidth,draft=false]{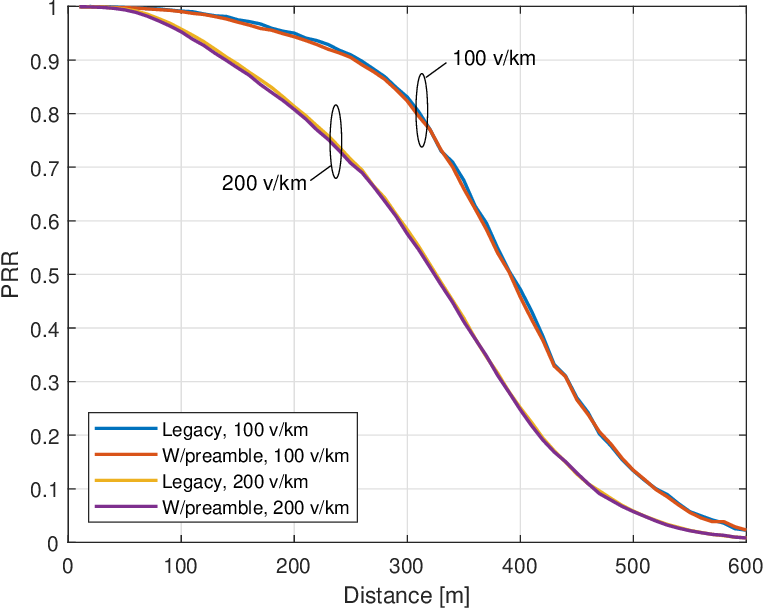}
	\caption{Performance of LTE-V2X only, with the preamble insertion and without (legacy). Highway scenario, 100 and 200~v/km.
	}
	\label{fig:simPRRvsDlteonly} \vskip -0.3cm
\end{figure} 

\begin{figure*} [t]
	\centering
	\subfigure[Packet reception ratio of IEEE 802.11p.]{
		\includegraphics[width=0.40\linewidth,draft=false]{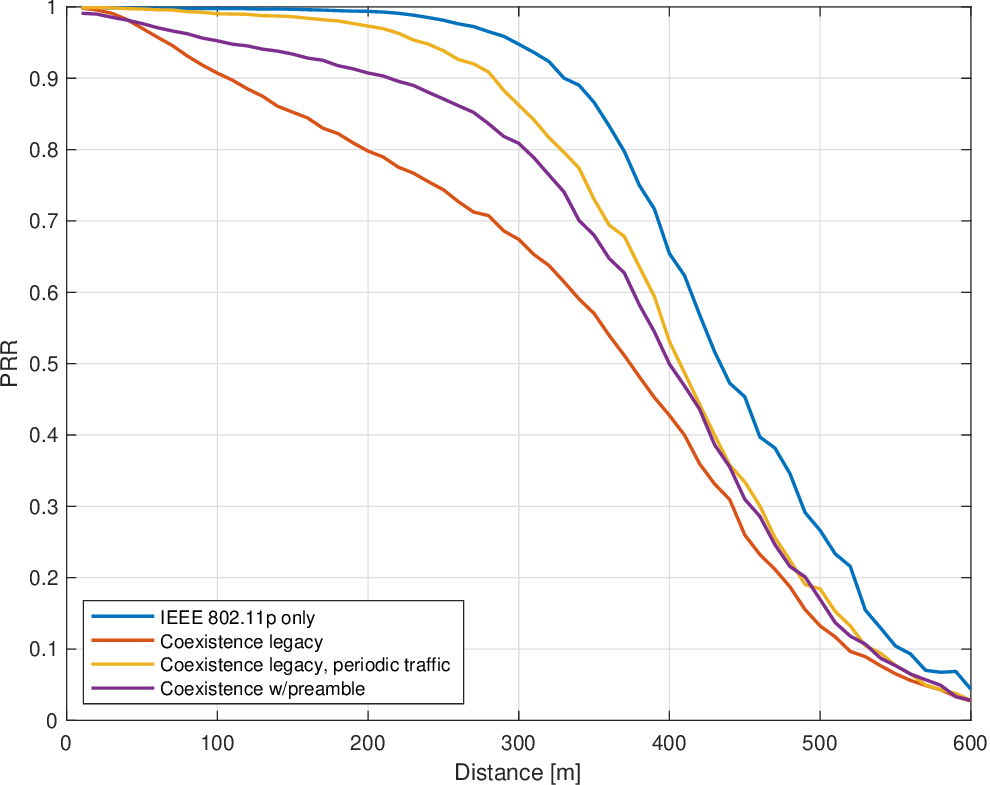}\label{fig:PRRa}}~~~~
	\subfigure[Packet reception ratio of LTE-V2X.]{
		\includegraphics[width=0.40\linewidth,draft=false]{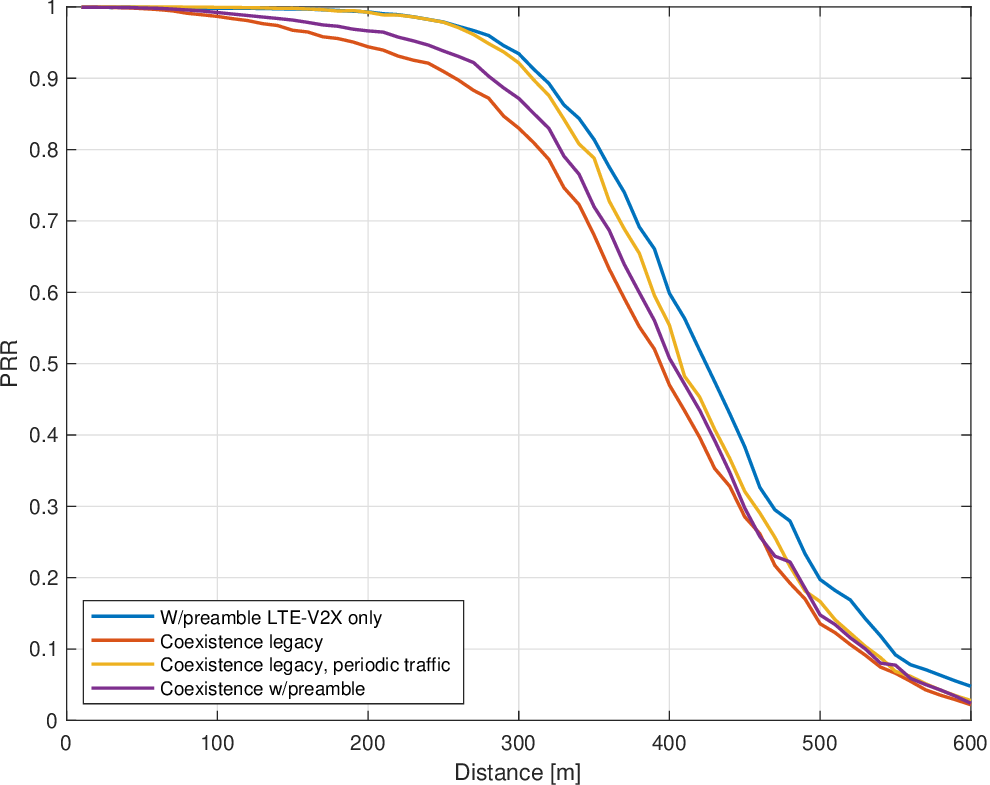}\label{fig:PRRb}}\\
\subfigure[Data age of IEEE 802.11p.]{
		\includegraphics[width=0.40\linewidth,draft=false]{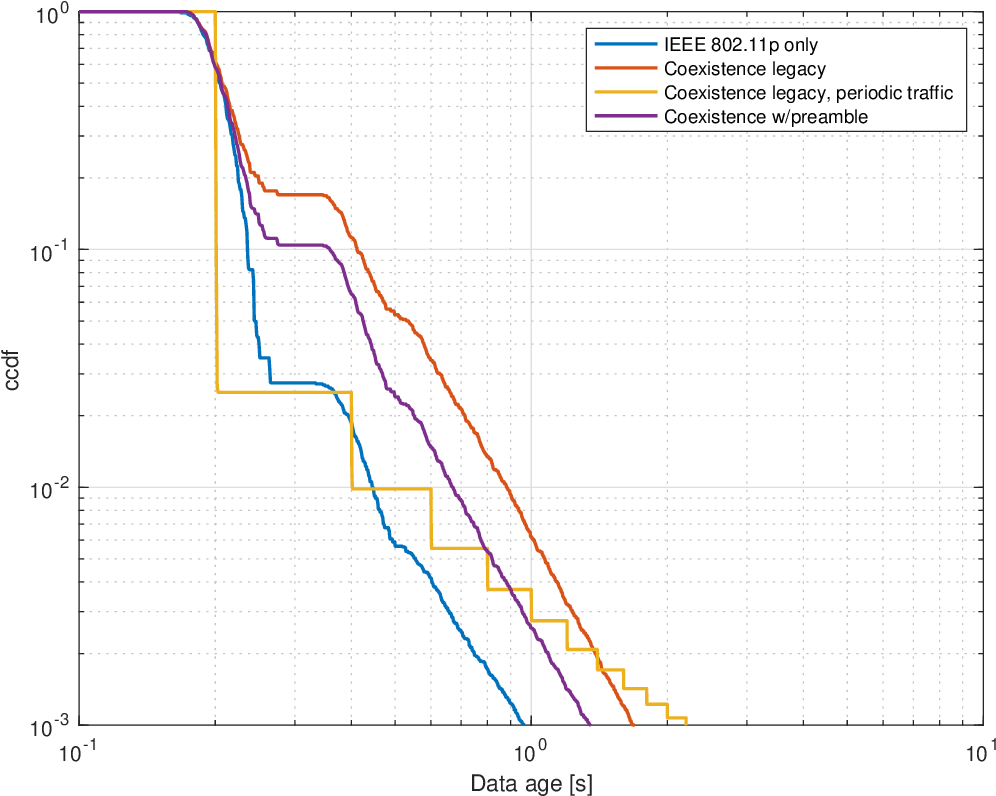}\label{fig:DAa}}~~~~
	\subfigure[Data age of LTE-V2X.]{
		\includegraphics[width=0.40\linewidth,draft=false]{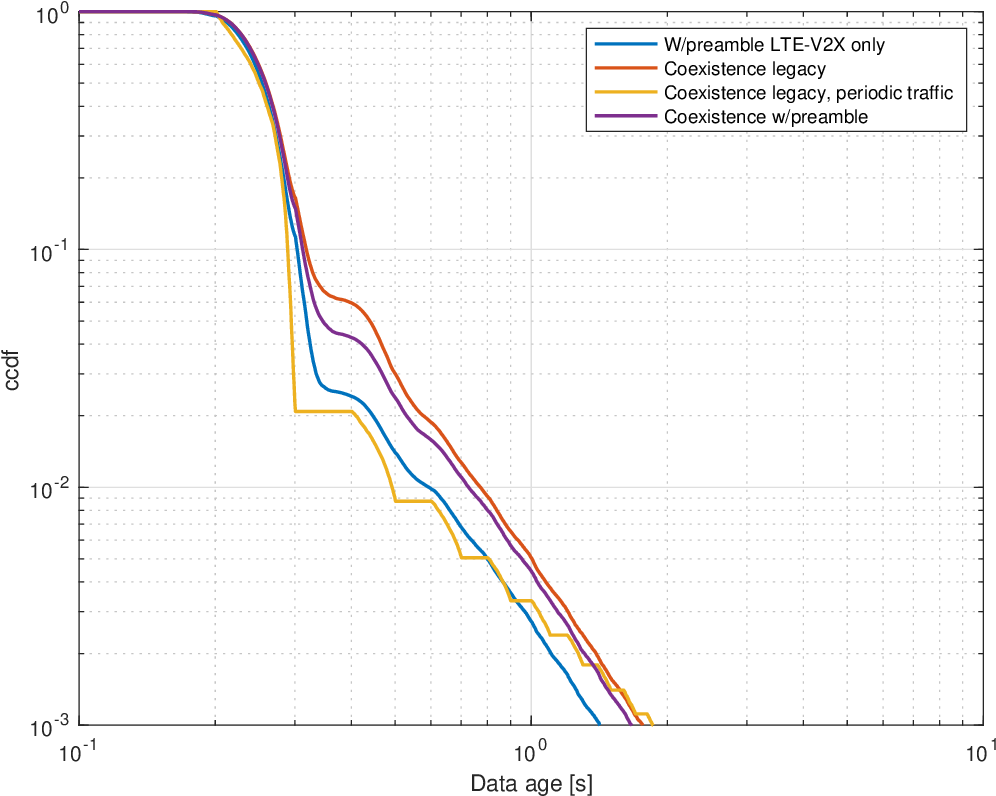}\label{fig:DAb}}
	\caption{Comparison between: (i) single technology with 50 v/km; (ii) coexistence of 50 IEEE~802.11p and 50~LTE-V2X v/km without any modifications; (iii) coexistence of 50 IEEE~802.11p and 50~LTE-V2X v/km without any modifications but with periodic traffic; (iv) coexistence of 50 IEEE~802.11p and 50~LTE-V2X v/km with the preamble insertion.}
	\label{fig:PRRDA}
	\vskip -0.3cm
\end{figure*}	

However, the power increase in the unused subchannels is very small, as it is slightly more than 4\% of the transmitted power in those actually used (slightly more than half of 13 OFDM symbols).  
The curves shown in Fig.~\ref{fig:simPRRvsDlteonly}, which refer to a scenario with 100 or 200 LTE-V2X stations per kilometer, both with and without preamble insertion, show that this effect is actually negligible.

\subsection{Performance of both technologies in terms of PRR and DA}\label{subsec:cochannel}

The impact of preamble insertion from the point of view of both technologies is shown in Fig.~\ref{fig:PRRDA}, which compares, in a scenario with 100 vehicles per technology per kilometer, the following cases: 
\begin{itemize}
    \item \textit{IEEE~802.11p/LTE-V2X only}: vehicles are equipped with the technology indicated; this case is used as a benchmark, without inter-technology interference; 
    \item \textit{Coexistence legacy:} IEEE 802.11p and LTE-V2X legacy stations share the channel; 
    \item \textit{Coexistence legacy, periodic traffic:} the two legacy technologies share the channel and the traffic generation is strictly periodic; in this case, to have a similar average number of packets per station per second, the generation interval of both technologies and the allocation period of LTE-V2X are set to 200~ms;
    \item \textit{Coexistence w/preamble:} the two technologies share the same channel and the preamble insertion is used.
\end{itemize} 
The case with periodic traffic is used to investigate what happens if the \ac{SB-SPS} of LTE-V2X is able to sense the use of the channel by IEEE~802.11p stations, which corresponds to the proposed inter-technology interference mitigation in \cite{BazZanSarMar:C20}. 

In particular, Figs.~\ref{fig:PRRa} and~\ref{fig:PRRb} provide the PRR varying the transmitter-receiver distance, with focus on IEEE~802.11p and LTE-V2X, respectively. By looking at Fig.~\ref{fig:PRRa} and comparing the curves with IEEE 802.11p alone and coexistence legacy, the presence of LTE-V2X is shown to significantly reduce the PRR. Both the use of a periodic generation of packets and the preamble insertion allow a significant mitigation of the inter-technology interference. Similar considerations can also be inferred from Fig.~\ref{fig:PRRb} when referring to LTE-V2X, although the negative impact of IEEE 802.11p on the PRR of LTE-V2X is smaller than in the reverse case. 

In Figs.~\ref{fig:DAa} and~\ref{fig:DAb}, the \ac{ccdf} of the DA is shown by referring to IEEE~802.11p and LTE-V2X, respectively. These plots confirm the conclusions derived from Figs.~\ref{fig:PRRa} and~\ref{fig:PRRb}, except for the case of periodic packet generation. In this case, the periodic transmissions increase the probability of consecutive collisions, implying a higher DA. In fact, if we look at a ccdf of 0.001 (DA occurring with probability 0.001 or less), we notice that the largest value is the one corresponding to periodic generation. Also from this perspective, the preamble insertion shows an improvement over legacy coexistence.

\input{TableNtx.tex}

\begin{figure*} [t]
	\centering
	\subfigure[Packet reception ratio of IEEE 802.11p.]{
		\includegraphics[width=0.40\linewidth,draft=false]{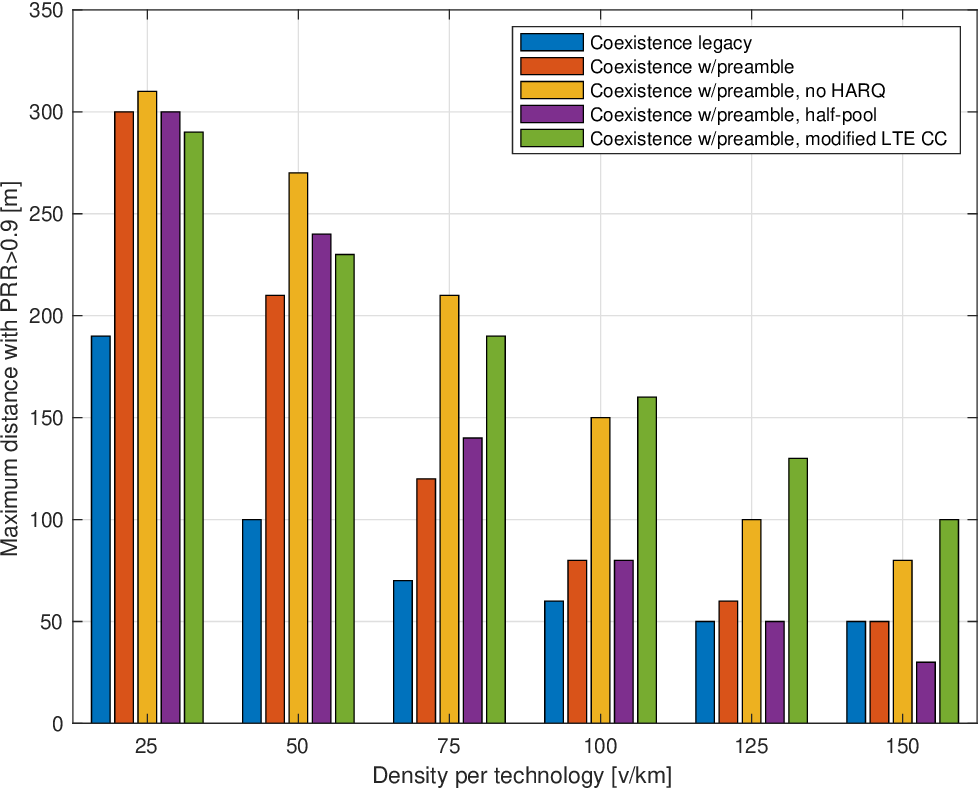}\label{fig:Hista}}~~~~
	\subfigure[Packet reception ratio of LTE-V2X.]{
		\includegraphics[width=0.40\linewidth,draft=false]{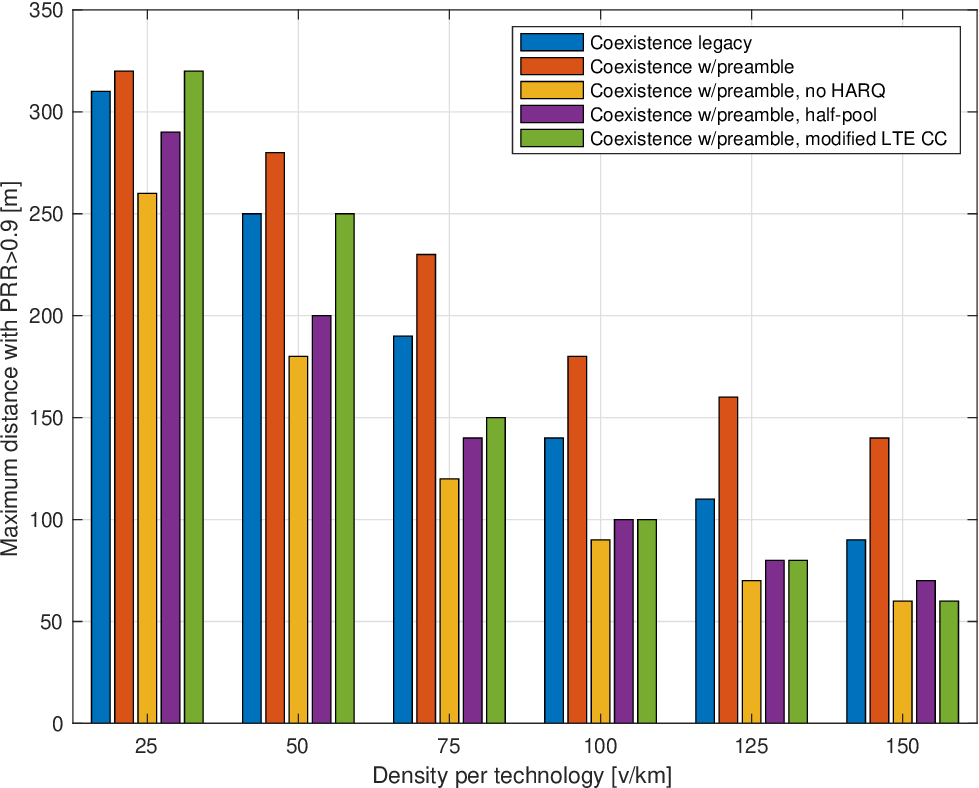}\label{fig:Histb}}\\
	\subfigure[Data age of IEEE 802.11p.]{
		\includegraphics[width=0.40\linewidth,draft=false]{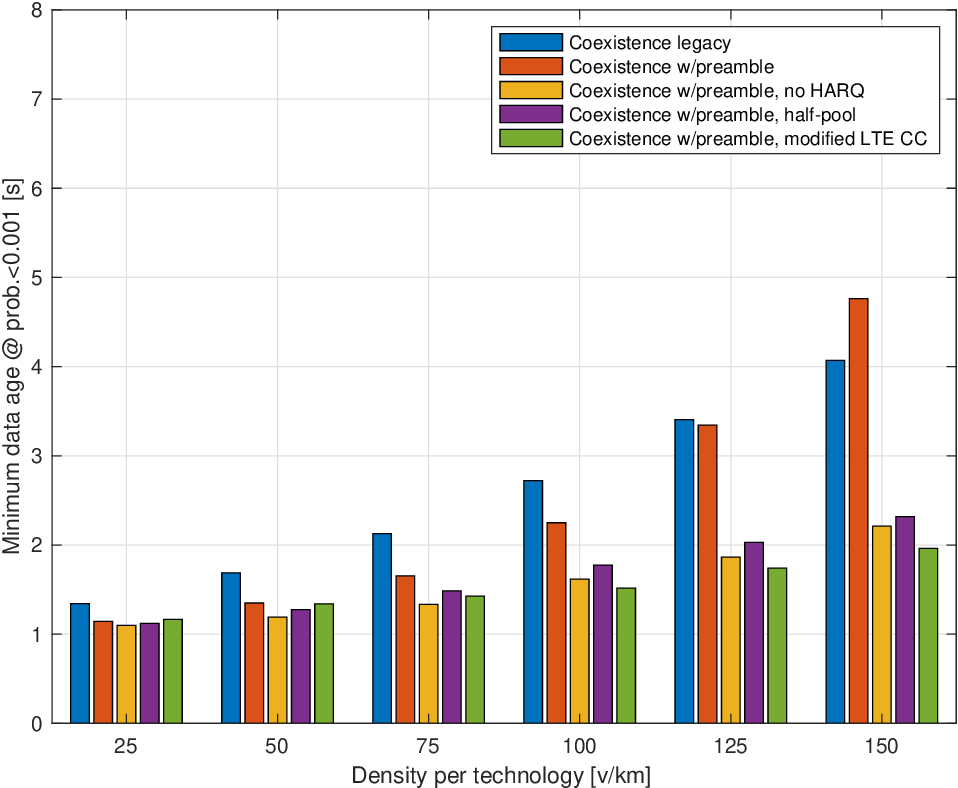}\label{fig:Histe}}~~~~
	\subfigure[Data age of LTE-V2X.]{
		\includegraphics[width=0.40\linewidth,draft=false]{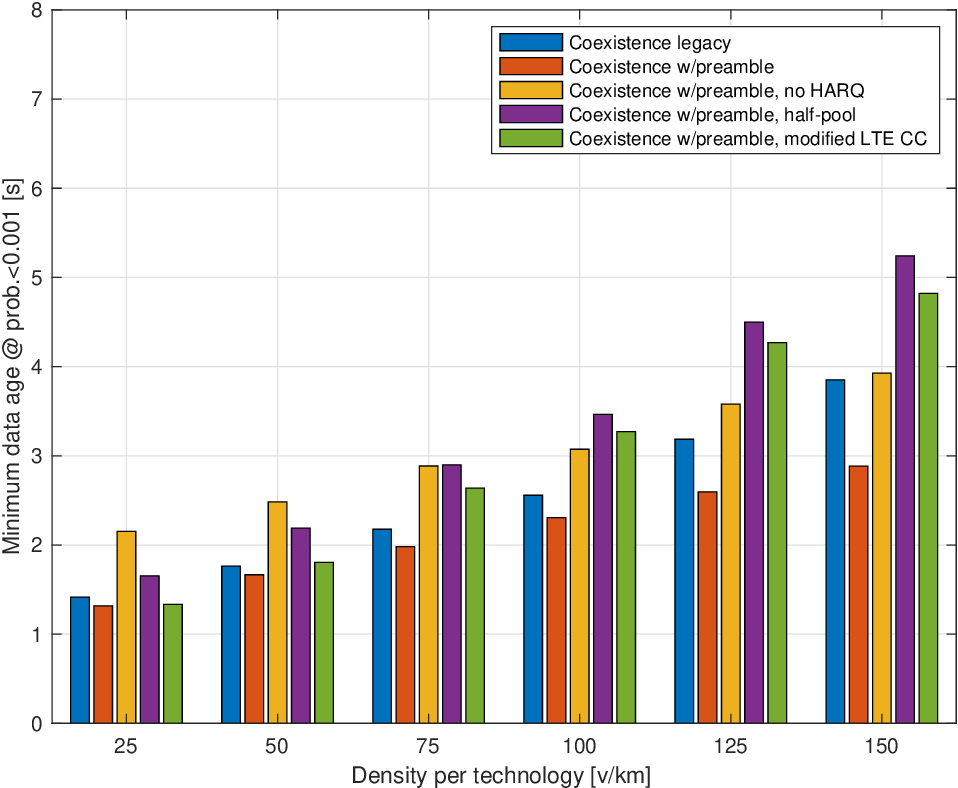}\label{fig:Histf}}\\
	\caption{Comparison of the coexistence of IEEE~802.11p and LTE-V2X vehicles, without any modifications or with preamble insertion, in the latter case without or with additional limitations.}
	\label{fig:Hist}\vskip -0.3cm
\end{figure*}

Similar results are also observed varying vehicle densities, not shown here for brevity reasons. A comparison of  coexistence with legacy and with preamble insertion for various densities can in any case be obtained by looking at the first two bars of the series shown in Fig.~\ref{fig:Hist} (i.e., blue and red bars). In Figs.~\ref{fig:Hista} and \ref{fig:Histb}, the maximum distance with PRR greater than 0.9 is shown by varying the density of the scenario; we can observe that the use of the preamble (red bars) guarantees performance higher than the legacy case (blue bars), with the exception of IEEE~802.11p with 150+150 v/km, where they behave similarly. Figs.~\ref{fig:Histe}, and~\ref{fig:Histf} show the minimum DA with probability less than 0.001. Again, the use of the preamble (red bars) leads to a lower DA for both technologies compared to legacy coexistence (blue bars), with the sole exception of IEEE~802.11p with 150+150~v/km, as explained in the next subsection.

\subsection{Considerations on the CBR}\label{Subsec:CBR}


In Table~\ref{Tab:Ntx}, for each density and simulated case, and for both technologies, the average number of messages generated and the average \ac{CBR} are reported. The CBR indicates the portion of resources currently estimated as being used by the generic node, with more details provided in Appendix~C. In the case of LTE-V2X, the average number of transmissions per packet is also shown in Table~\ref{Tab:Ntx}.

If attention is focused on high density scenarios (i.e., 150+150~v/km), it can be observed that in the case of  preamble insertion, \ac{CC} reduces the average messages generated in IEEE~802.11p to less than 3 per station per second; this eventually causes the \ac{DA} to increase, as observed in  Fig.~\ref{fig:Histe}.

\section{Impact in congested scenarios}\label{Sec:congested}

As already noted with reference to the first two bars of the series in  Fig.~\ref{fig:Hist} and deepened in Section~\ref{Subsec:CBR}, when vehicle density increases, LTE-V2X stations tend to use most of the resources, the preamble tends to be insufficient to mitigate the coexistence issues, and eventually the IEEE 802.11p traffic is reduced by its \ac{CC} mechanism. 

For this reason, in this section approaches added to preamble insertion are considered and compared to prevent LTE-V2X stations from taking most of the channel in congested scenarios. Results are obtained via simulations, with the settings summarized in Table~\ref{Tab:Settings} and in terms of PRR and DA.

The implemented approaches, which are all in addition to preamble insertion, are as follows:
\begin{itemize}
    \item \textit{No \ac{HARQ}:} in LTE-V2X, the use of blind retransmissions is inhibited; this prevents LTE-V2X stations from performing two transmissions per packet, thus making the traffic generated by the two technologies similar;
    \item \textit{Half pool:} LTE-V2X stations can only use part of the subframes; this approach, also discussed in \cite{ETSI_TR_103_766}, is allowed in LTE-V2X thanks to the concept of pool of accessible resources; a pool of 25 subframes every 50~ms is assumed (corresponding to 50\%);\footnote{The so-called \textit{method C} in \cite{ETSI_TR_103_766} assumes that the LTE-V2X nodes somehow estimate locally the proportion of stations equipped with each technology and consistently set the pool; therefore, setting the pool to 50\% is the ideal output of that process in the scenario of this paper; this method is the only one in \cite{ETSI_TR_103_766} fully compatible with current deployments and it has been  shown to provide the best performance among those not requiring a centralized coordination; comparison of the methods is given in \cite{ETSI_TR_103_766} and is outside the scope of the current work.}
    \item \textit{Modified LTE-V2X CC:} the \ac{CC} defined in \cite{ETSI_103_574} is modified to reduce the use of the channel by LTE-V2X; in particular, all thresholds that control channel occupation are halved, as detailed in Appendix~C. We adopted exactly half for all thresholds as a simple solution. The optimization of the \ac{CC} mechanism is beyond the scope of the present paper and left for future work.
\end{itemize} 

The impact of the three approaches in terms of generated messages and channel occupation is observable in Table~\ref{Tab:Ntx}, while the PRR and DA are shown in Fig.~\ref{fig:Hist}. 

As for \textit{no HARQ}, Table~\ref{Tab:Ntx} shows that it reduces the number of transmissions per message in LTE-V2X to one and this allows IEEE 802.11p to maintain about 4.9 average messages per station per second even with 150+150~v/km. This also allows a significant improvement of the PRR of IEEE 802.11p, as shown in Fig.~\ref{fig:Hista}. At the same time, LTE-V2X cannot take advantage of one of its features, which implies a significant loss of performance even when it is not needed.


Moving on to the \textit{half pool}, 
it allows both technologies to maintain the same average messages per station per second even with the highest vehicle density (see Table~\ref{Tab:Ntx}). However, as observable in Figs.~\ref{fig:Hista}-\ref{fig:Histb}, the PRR of IEEE 802.11p improves only when the vehicle density is low. In addition, the PRR of LTE-V2X is always less than the legacy case and the one with the preamble insertion.  This is due to the fact that although LTE-V2X stations leave half of the subframes free to be used by IEEE 802.11p, the average number of LTE-V2X transmissions performed within the subframes of the allowed pool is doubled, with a negative effect in both technologies.

The last solution, which is the \textit{modified LTE-V2X CC}, shows the best performance for IEEE~802.11p under high density conditions. The performance reduction of LTE-V2X, compared to the legacy case or the one with preamble insertion alone, is similar to the other two approaches, with a DA which is better than \textit{no HARQ} for low vehicle density and of \textit{half pool} in all the cases.

In general, it can be observed that an improvement in one technology corresponds to a loss of performance in the other. However, adopting the preamble insertion with a modification of the \ac{CC} mechanisms of LTE-V2X allows in low traffic conditions to achieve near-maximum performance in both technologies, and in conditions of increased traffic provides a better balance between the performances of the two technologies.

\section{Conclusion}\label{sec:conclusion}

In this work, focusing on a scenario where IEEE 802.11p and sidelink LTE-V2X Mode~4 coexist in the same geographical area and in the same channel, we have studied the insertion of the IEEE~802.11p preamble at the beginning of the LTE-V2X transmissions to mitigate the mutual interference between technologies. This preamble insertion does not require modifications to IEEE~802.11p and is therefore compatible with current deployments in Europe. In addition, it implies only minor changes to LTE-V2X. To access the impact of this proposal on the performance of the two technologies, we first obtained an analytical model in free-flow scenarios. The result is a closed-form expression that demonstrates the significant reduction of collisions due to inter-technology interference. The study has been then extended to denser scenarios through the use of an open-source simulator, confirming the effectiveness of the proposal. Overall, the solution based on preamble insertion reduces performance losses caused by co-channel coexistence, without affecting the performance of individual technologies in areas or time intervals without coexistence. It has also been shown that mitigation loses effectiveness in particularly high traffic conditions. For such situation, three different approaches have been compared in addition to the preamble insertion, among which the one that acts on the congestion control algorithm of LTE-V2X has proven to be the most effective.

\section*{Appendix A: Derivation of \eqref{eq:PRRsubseqApproxCF}}

Since the main interferer is in this case outside the protected range, the \ac{PRP} is 
\small
\begin{align}\label{eq:A1}
    \PRRbusy = & \frac{1}{1-\Psensed}\int_{\mathbb{R}-\{\mathbb{S}\}}\Ppos(x) \fPER{\frac{\Pru}{\Pn+\Ptxlte\Gt\Gr/\Loss(x)}}dx
\end{align}
\normalsize
where $\mathbb{S}$ is the \textit{protected area}, i.e., from $\du-\dmaxx$ to $\du+\dmaxx$ and $\mathbb{R}-\{\mathbb{S}\}$ indicates from $-\infty$ to $\infty$ except $\mathbb{S}$. 
The term before the integral in \eqref{eq:A1} is a normalization due to the assumption of having the LTE-V2X interferer outside the protected area and the integrating represents the conditioned  \ac{PRP} for the possible positions of the interferer.

Approximating, as indicated in Section~\ref{subsec:modelassumptions}, the $\fPER{\SINRu}$ to a threshold curve, the SINR threshold corresponds to a maximum interference value above which the packet is lost, which in turn corresponds to the minimum distance for the interferer $\dImin$. Given this approximation, \eqref{eq:A1} can be written as 
\begin{align}\label{eq:A2}
    \PRRbusy = & \frac{1}{1-\Psensed}\left[\int_{-\infty}^{-\max\{\dImin,(\dmaxx-\du)\}}\Ppos(x)dx \right.\nonumber \\ &\left.+ \int_{\max\{\dImin,(\dmaxx+\du)\}}^{\infty}\Ppos(x)dx\right]
\end{align}
which leads to \eqref{eq:PRRsubseqApproxCF}.

\section*{Appendix B: Derivation of \eqref{eq:PRRsubseqApprox} and \eqref{eq:prrunprotectedCF}}

In the case the current TTI is sensed idle by the IEEE~802.11p transmitter and the transmission ends in the subsequent TTI, the \ac{PRP} is equal to 
\begin{align}\label{eq:B1}
    \PRRsubseq=&
    \frac{1}{1-\Psensed}\int_{0}^{1}\left[\int_{-\infty}^{\infty}\Ppos(y)\right.\nonumber\\&\left.\cdot\left[
\int_{\mathbb{R}-\{\mathbb{S}\}} \Ppos(x) 
f_\text{xy}(x,y,\tau)
 dx\right] dy\right]d\tau  \;. 
\end{align}
where 
\small
\begin{align}
    f_\text{xy}(x,y,\tau)=\fPER{\frac{\Pru}{\Pn+(1-\tau)\frac{\Ptxlte\Gt\Gr}{\Loss(x)}+\tau\frac{\Ptxlte\Gt\Gr}{\Loss(y)}}}
\end{align}
\normalsize
and the variable $\tau$ indicates the portion of the packet transmitted during the subsequent TTI.
Assuming that the interference is entirely caused by the LTE-V2X transmission that overlaps more with the reference transmission (i.e., that in the current TTI if $\tau<0.5$ and that in the subsequent TTI if $\tau\geq0.5$), \eqref{eq:B1} can be approximated as 
\small
\begin{align}\label{eq:B2}
    \PRRsubseq=&
    \frac{1/2}{1-\Psensed} \left[
\int_{\mathbb{R}-\{\mathbb{S}\}} \Ppos(x) 
\fPER{\frac{\Pru}{\Pn+\Ptxlte\Gt\Gr/\Loss(x)}}
 dx\right] \nonumber\\
 &+\frac{1}{2} \int_{-\infty}^{\infty}\Ppos(y)
\fPER{\frac{\Pru}{\Pn+\Ptxlte\Gt\Gr/\Loss(y)}}
dy \nonumber \\ =& \frac{\PRRbusy}{2} + \frac{ \PRRunprotected}{2}
\end{align}
\normalsize
where
\begin{align}\label{eq:B3}
\PRRunprotected=\int_{-\infty}^{\infty}\Ppos(y)
\fPER{\frac{\Pru}{\Pn+\Ptxlte\Gt\Gr/\Loss(y)}}
dy\;.
\end{align}
Directly, \eqref{eq:B2} corresponds to \eqref{eq:PRRsubseqApprox}. Furthermore, from \eqref{eq:B3}, applying the threshold model for the \ac{PER} vs. SINR, we obtain 
\begin{align}\label{eq:B4}
\PRRunprotected=\int_{-\infty}^{-\dImin}\Ppos(y)dy+\int_{\dImin}^{\infty}\Ppos(y)dy
\end{align}
which leads  to \eqref{eq:prrunprotectedCF}.

\section*{Appendix C: Congestion Control}

In Section~\ref{sec:resSystem}, the \ac{CC} mechanisms described in \cite{ETSI_EN_302_663} and \cite{ETSI_103_574} are used for IEEE~802.11p and LTE-V2X, respectively.

\subsubsection{IEEE 802.11p} \ac{DCC} detailed for ITS-G5 in \cite{ETSI_EN_302_663}; each station measures the portion of time during which the received power is greater than -85~dBm, within intervals of 100~ms, called \ac{CBR} and here denoted as $\deltacbrp$; denoting as $t_\text{g}$ the generation interval between one packet and the next as indicated by the higher layers from \cite{3GPP_EN_302_637_2}, the interval used between consecutive packets is calculated as $t_\Delta=\max\{t_\text{g},\min\{1,t_\text{g}\cdot4000\cdot\frac{\deltacbrp-0.62}{\deltacbrp}\}\}$, where $\min\{x,y\}$ is a function that returns the minimum value between $x$ and $y$. If the CBR goes above 0.62, the generation interval is increased in order to reduce the channel occupation.

\subsubsection{LTE-V2X} \ac{CC} detailed in \cite{ETSI_103_574}; each station measures every 100~ms the portion of subchannels with received power greater than -94~dBm, called \ac{CBR} and here denoted as $\deltacbrlte$; based on this and using the settings for the \acp{CAM}, the average number of subchannels that can be used per second by the station, called \ac{CR} and here denoted as $\rhoCR$, is constrained to $\rhoCR<0.03$ if $0.3<\deltacbrlte\leq0.65$, $\rhoCR<0.006$ if $0.65<\deltacbrlte\leq 0.8$, and $\rhoCR<0.003$ if $\deltacbrlte>0.8$. The way to reduce the load is not specified in the standards and various options are possible (e.g., varying the \ac{MCS} or power). In the simulator, we assume that in order to comply with the constrain on the $\rhoCR$, first blind retransmissions are avoided, and then, if not sufficient, the same equation used for IEEE~802.11p is applied to the generation interval.

\subsubsection{LTE-V2X with modified CC} same algorithm with stricter constraints; in particular, all numbers are halved, which means that $\rhoCR<0.015$ if $0.15<\deltacbrlte\leq0.325$, $\rhoCR<0.003$ if $0.325<\deltacbrlte\leq 0.4$, and $\rhoCR<0.0015$ if $\deltacbrlte>0.4$.

\bibliographystyle{IEEEtran}
\bibliography{biblioSelf,biblioOthers,biblioStandards}

\end{document}

%% file: Definitions.tex
\newcommand{\du}{d_\text{u}}
\newcommand{\dIfirst}{x}
\newcommand{\dIsecond}{y}
\newcommand{\Ptxp}{P_\text{11p}}
\newcommand{\Ptxlte}{P_\text{LTE}}
\newcommand{\Gt}{G_\text{t}}
\newcommand{\Gr}{G_\text{r}}
\newcommand{\Pru}{P_\text{R}}

\newcommand{\Prminx}{{{P_{\text{R},x}^{*}}}}

\newcommand{\Pn}{P_\text{N}}
\newcommand{\PI}{P_\text{I}}

\newcommand{\PRR}{\mathbb{P}_\text{PR}}
\newcommand{\PRRcurrent}{\mathbb{P}_\text{PR$|$c}}
\newcommand{\PRRsubseq}{\mathbb{P}_\text{PR$|$sq}}
\newcommand{\PRRbusy}{\mathbb{P}_\text{PR$|$busy}}
\newcommand{\PRRunprotected}{\mathbb{P}_\text{PR$|$unpr}}
\newcommand{\fPER}[1]{f_\text{PER}\left(#1\right)}
\newcommand{\SINRu}{\gamma}
\newcommand{\SINRt}{\underline{\gamma}}

\newcommand{\Loss}{L}
\newcommand{\Pwithin}{\mathbb{P}_\text{c}}
\newcommand{\Pcross}{\mathbb{P}_\text{sq}}
\newcommand{\Psensed}{\mathbb{P}_\text{busy}}
\newcommand{\Pnotswithin}{{\mathbb{P}_\text{c-idle}}}
\newcommand{\Pnotscross}{{\mathbb{P}_\text{sq-idle}}}
\newcommand{\Ppos}{\mathbb{P}_\text{d}}
\newcommand{\tTTI}{t_\text{TTI}}
\newcommand{\tPack}{t_\text{pck}}

\newcommand{\dmaxx}{d^*_x}
\newcommand{\dImin}{\underline{d_\text{i}}}

\newcommand{\averageLTEtx}{\lambda}
\newcommand{\averageLTEtxN}{\lambda_\text{TTI}}
\newcommand{\NttiSec}{N_\text{TTI}}
\newcommand{\deltacbrp}{\delta_\text{CBR-11p}}
\newcommand{\deltacbrlte}{\delta_\text{CBR-LTE}}
\newcommand{\rhoCR}{\rho_\text{CR-LTE}}

%% file: Acronyms.tex
\begin{acronym} 
\acro{3GPP}{Third Generation Partnership Project}
\acro{5GAA}{5G Automotive Association}
\acro{AGC}{automatic gain control}
\acro{AIFS}{arbitration inter-frame space}
\acro{AWGN}{additive white Gaussian noise}
\acro{BEP}{beacon error probability}
\acro{BP}{beacon periodicity}
\acro{BSM}{basic safety message}
\acro{BSS}{basic service set}
\acro{ccdf}{complementary cumulative distribution function}
\acro{cdf}{cumulative distribution function}
\acro{CA}{collision avoidance}
\acro{CC}{congestion control}
\acro{CCA}{clear channel assessment}
\acro{CAM}{cooperative awareness message}
\acro{CAMs}{cooperative awareness messages}
\acro{CAV}{connected and automated vehicle}
\acro{CBR}{channel busy ratio}
\acro{C-ITS}{cooperative-intelligent transport systems}
\acro{CDMA}{code division multiple access}
\acro{CR}{channel occupation ratio}
\acro{CSD}{cyclic shift diversity}
\acro{CSI}{channel state information}
\acro{CSMA/CA}{carrier sensing multiple access with collision avoidance}
\acro{CSS}{chirp spread spectrum}
\acro{C-V2X}{cellular-\ac{V2X}}
\acro{D2D}{device-to-device}
\acro{DA}{data age}
\acro{DCC}{distributed congestion control}
\acro{DCF}{distributed coordination function}
\acro{DCM}{Dual carrier modulation}
\acro{DENM}{decentralized environmental notification message}
\acro{DMRS}{demodulation reference signal}
\acro{DSRC}{dedicated short range communication}
\acro{EDCA}{enhanced distributed coordination access}  
\acro{eNB}{evolved NodeB}
\acro{FCD}{floating car data}
\acro{FD}{full duplex}
\acro{FDD}{frequency division duplex}
\acro{FEC}{forward error correction}
\acro{FHSS}{frequency hopping spread spectrum}
\acro{GNSS}{global navigation satellite system}
\acro{GPS}{global positioning system}
\acro{HARQ}{hybrid automatic request}
\acro{HD}{half duplex}
\acro{IBE}{in-band emission}
\acro{IPG}{Inter-packet gap}
\acro{ISM}{industrial, scientific and medical}
\acro{ITS}{intelligent transport system}
\acro{i.i.d.}{independent identically distributed}
\acro{KPI}{key performance indicator}  
\acro{LDPC}{low density parity check}
\acro{LOS}{line-of-sight}
\acro{LTE}{long term evolution}  
\acro{LTE-D2D}{\ac{LTE} with \ac{D2D} communications}  
\acro{LTE-V2V}{\ac{LTE}-vehicle-to-vehicle}
\acro{LTE-V2X}{\ac{LTE}-\ac{V2X}}
\acro{LTE-D2D}{long term evolution with device to device communications}  
\acro{MAC}{medium access control}
\acro{MCS}{modulation and coding scheme}
\acro{MIMO}{multiple input multiple output}
\acro{MRD}{maximum reuse distance}
\acro{NGV}{Next Generation V2X}
\acro{NHTSA}{National Highway Traffic Safety Administration}
\acro{NLOS}{non-line-of-sight}
\acro{NR}{new radio}
\acro{NR-V2X}{\ac{NR}-\ac{V2X}}
\acro{OBU}{on board unit}
\acro{OCB}{outside of the context of a \acl{BSS}}
\acro{OFDM}{orthogonal frequency division multiplexing}
\acro{OFDMA}{orthogonal frequency division multiple access}
\acro{pdf}{probability density function}
\acro{PAPR}{peak to average power ratio}
\acro{PD}{packet delay}
\acro{PEP}{pairwise error probability} 
\acro{PER}{packet error rate}
\acro{PHY}{physical}
\acro{PL}{path loss}
\acro{PPP}{Poisson point
process}
\acro{PRB}{physical resource block}
\acro{ProSe}{proximity-based services}
\acro{PRP}{packet reception probability}
\acro{PRR}{packet reception ratio}
\acro{QAM}{quadrature amplitude modulation}
\acro{QoS}{quality of service}
\acro{RR}{radio resource}
\acro{RSU}{road side unit} 
\acro{SB-SPS}{sensing-based semi-persistent scheduling}
\acro{SC-FDMA}{single carrier frequency division multiple access}
\acro{SCI}{sidelink control information}
\acro{SHINE}{simulation platform for heterogeneous interworking networks}
\acro{SINR}{signal to noise and interference ratio}
\acro{SNR}{signal to noise ratio}
\acro{SRS}{sounding reference signal}
\acro{STBC}{space time block codes}
\acro{SV}{smart vehicle}
\acro{TB}{transport block}
\acro{TDD}{time division duplex}
\acro{TTI}{transmission time interval}
\acro{UE}{user equipment}
\acro{URLLC}{ultra reliable and low latency communications}
\acro{UTDOA}{uplink time difference of arrival}
\acro{V2C}{vehicle-to-cellular} 
\acro{V2N}{vehicle-to-network}
\acro{V2I}{vehicle-to-infrastructure} 
\acro{V2P}{vehicle-to-pedestrian}
\acro{V2R}{vehicle-to-roadside}
\acro{V2V}{vehicle-to-vehicle} 
\acro{V2X}{vehicle-to-everything} 
\acro{WAVE}{wireless access in vehicular environment}
\acro{WCDMA}{wideband code division multiple access}
\end{acronym}

%% file: TableSettings.tex
\begin{table}
\caption{Main adopted parameters and settings.
\vspace{-2mm}
\label{Tab:Settings}}
\scriptsize
\centering
\begin{tabular}{p{3cm}p{4.7cm}}
\hline \hline
\textbf{\textit{Free-flow scenario}} & \\
Scenario & Highway, approximated as 1-D\\
Density & Variable \\
Average transmissions & Variable \\ \hline
\textbf{\textit{Denser scenarios}} & \\
Scenario & 3+3 lanes highway, 2~km straight road \cite{3GPP_TR_36_885}\\
Density & Variable \\
Mobility & Gaussian distributed speed, with average 70\;km/h and std. dev. 7\;km/h \\
Packet periodicity & Following CAM rules \cite{3GPP_EN_302_637_2} \\
\hline \hline
\textbf{\textit{Common settings}} & \\
Channels & ITS bands at 5.9 GHz \\
Bandwidth & 10 MHz \\
Transmission power density & 13~dBm/MHz \\
Antenna gain (tx and rx)  & 3 dB \cite{BazZanSarMar:C20} \\
Noise figure & 6~dB \cite{BazZanSarMar:C20} \\
Propagation model & WINNER+, Scenario B1, line-of-sight \cite{3GPP_TR_36_885}\\
Shadowing & Variance 3 dB, decorr. dist. 25~m \cite{3GPP_TR_36_885}\\
Packet size & 350~B \cite{Car2Car_CAMstats} \\ 
\hline
\textbf{\textit{IEEE~802.11p}} & \\
MCS & 2 (QPSK, 1/2), PER=0.5@SINR=1.02\;dB \\
Duration of the initial space & 110\;$\mu$s \cite{ETSI_TS_102_636} \\
Random backoff & $[0\div 15] \cdot 13$\;$\mu$s \cite{ETSI_TS_102_636} \\ 
Carrier sense threshold & -65 dBm \\
Preamble detection threshold & -98.8 dBm (see Section~\ref{Subsec:freeflowresults}) \\ 
Congestion control & ETSI DCC \cite{ETSI_EN_302_663} (see Appendix~C) \\
\hline
\textbf{\textit{Sidelink LTE-V2X Mode 4}} & \\
MCS & 11 (16-QAM, 0.41), PER=0.5@SINR=5.15\;dB \\
Subchannel size & 10 resource block pairs \cite{ETSI_TS_103_613} \\
Number of subchannels & 5 \\
Subchannels per packet & 2 \\
Configuration & Adjacent \\
Keep probability & 0.5 \\
Allocation periodicity & 100 ms \cite{BarMasMarSarBaz:J21} \\
Subchannel sensing threshold & -110~dBm \\
Congestion control & ETSI CC for LTE-V2X \cite{ETSI_103_574} (see Appendix~C) \\
HARQ & Blind retransmission if CC allows \\
\hline \hline
\end{tabular}
\end{table}

%% file: TableNtx.tex
\begin{table*}
\caption{Average number of messages (Msg) and average CBR (CBR) for IEEE 802.11p and LTE-V2X. In LTE-V2X, the average number of transmissions (Ntx) per packet is also shown.
\vspace{-2mm}
\label{Tab:Ntx}}
\footnotesize
\centering
\begin{tabular}{p{3.8cm}|p{0.4cm}p{0.5cm}|p{0.4cm}p{0.4cm}p{0.5cm}|p{0.4cm}p{0.5cm}|p{0.4cm}p{0.4cm}p{0.5cm}|p{0.4cm}p{0.5cm}|p{0.4cm}p{0.4cm}p{0.5cm}}
\hline \hline
& \multicolumn{5}{c|}{50+50 v/km}& \multicolumn{5}{c|}{100+100 v/km}& \multicolumn{5}{c}{150+150 v/km} \\
& \multicolumn{2}{c|}{IEEE 802.11p} & \multicolumn{3}{c|}{LTE-V2X}& \multicolumn{2}{c|}{IEEE 802.11p} & \multicolumn{3}{c|}{LTE-V2X}& \multicolumn{2}{c|}{IEEE 802.11p} & \multicolumn{3}{c}{LTE-V2X}\\
\textbf{\textit{Case}} & Msg & CBR & Msg & Ntx & CBR & Msg & CBR & Msg & Ntx & CBR & Msg & CBR & Msg & Ntx & CBR \\ \hline
IEEE 802.11p only &4.87	&0.055&&& &4.84	&0.107&&& & 4.89 & 0.166 & & & \\
LTE-V2X only (w/preamble) &&&4.86	&2	&0.17 &&&4.83&	2&	0.316 & & &  4.88&	2	&0.446\\ \hline
Coexistence legacy &4.88	&0.192&4.79	&2&	0.3 &4.89	&0.358&4.81	&2	&0.52 & 4.89 & 0.496 & 4.85 & 1.99 & 0.684\\
Legacy w/ periodic generation &5	&0.21&5	&2&	0.343 &5	&0.405&5&	2	&0.617 & 5 & 0.532 & 5 & 1.87 &	0.770 \\
W/preamble &4.88	&0.401&4.79	&2&	0.322 &4.61	&0.635&4.81	&2&	0.542 & 2.73&	0.781 & 4.85&	2	&0.626 \\ \hline
W/preamble, no HARQ &4.88&	0.245&4.79&	1&	0.242 &4.89&	0.426&4.81&	1	&0.438 & 4.87	&0.568 & 4.85&	1	&0.598 \\
W/preamble, half pool &4.88	&0.343&4.79	&1.97&	0.308 &4.89	&0.491&4.81	&1.98	&0.516 & 4.89	&0.579 &4.85&	1.98&	0.662 \\
W/preamble, modified LTE CC &4.88&0.351&4.74&	1.72&	0.294 &4.89	&0.399&4.15	&1.11&	0.422 & 4.89&	0.493& 3.81&	1.02&	0.564 \\
\hline \hline
\end{tabular}
\end{table*}

%% file: main.bbl
\begin{thebibliography}{10}
\providecommand{\url}[1]{#1}
\csname url@samestyle\endcsname
\providecommand{\newblock}{\relax}
\providecommand{\bibinfo}[2]{#2}
\providecommand{\BIBentrySTDinterwordspacing}{\spaceskip=0pt\relax}
\providecommand{\BIBentryALTinterwordstretchfactor}{4}
\providecommand{\BIBentryALTinterwordspacing}{\spaceskip=\fontdimen2\font plus
\BIBentryALTinterwordstretchfactor\fontdimen3\font minus
  \fontdimen4\font\relax}
\providecommand{\BIBforeignlanguage}[2]{{%
\expandafter\ifx\csname l@#1\endcsname\relax
\typeout{** WARNING: IEEEtran.bst: No hyphenation pattern has been}%
\typeout{** loaded for the language `#1'. Using the pattern for}%
\typeout{** the default language instead.}%
\else
\language=\csname l@#1\endcsname
\fi
#2}}
\providecommand{\BIBdecl}{\relax}
\BIBdecl

\bibitem{6957143}
F.~M. {Abinader}, E.~P.~L. {Almeida}, F.~S. {Chaves}, A.~M. {Cavalcante}, R.~D.
  {Vieira}, R.~C.~D. {Paiva}, A.~M. {Sobrinho}, S.~{Choudhury}, E.~{Tuomaala},
  K.~{Doppler}, and V.~A. {Sousa}, ``Enabling the coexistence of {LTE} and
  {Wi-Fi} in unlicensed bands,'' \emph{IEEE Communications Magazine}, vol.~52,
  no.~11, pp. 54--61, 2014.

\bibitem{8806680}
B.~{Mafakheri}, L.~{Goratti}, R.~{Abbas}, S.~{Reisenfeld}, and R.~{Riggio},
  ``{LTE/Wi-Fi} coordination in unlicensed bands: An {SD-RAN} approach,'' in
  \emph{2019 IEEE Conference on Network Softwarization (NetSoft)}, 2019.

\bibitem{BazZanSarMar:C20}
A.~{Bazzi}, A.~{Zanella}, I.~{Sarris}, and V.~{Martinez}, ``Co-channel
  coexistence: Let {ITS-G5} and sidelink {C-V2X} make peace,'' in \emph{IEEE
  ICMIM, 2020}, pp. 1--4.

\bibitem{roux2020performance}
P.~Roux and V.~Mannoni, ``{P}erformance {E}valuation for {C}o-channel
  {C}oexistence {B}etween {ITS-G5} and {LTE-V2X},'' in \emph{IEEE VTC Fall
  2020}.

\bibitem{9644653}
M.~A. Ruder, M.~Papaleo, S.~Stefanatos, T.~V. Nguyen, and S.~Patil, ``On the
  coexistence between {LTE-V2X} sidelink and {ITS-G5},'' in \emph{2021 IEEE
  Vehicular Networking Conference (VNC)}, 2021, pp. 162--169.

\bibitem{ETSI_TR_103_766}
``Intelligent transport systems ({ITS}); pre-standardization study on
  co-channel co-existence between {IEEE}- and {3GPP}-based its technologies in
  the 5 855 {MHz}-5 925 {MHz} band,'' \emph{ETSI TR 103 766 v1.1.1}, Sept.
  2021.

\bibitem{IEEE80211_2020}
``{IEEE} 802.11-2020 - {IEEE} standard for information technology -
  telecommunications and information exchange between systems - local and
  metropolitan area networks--specific requirements - part 11: Wireless lan
  medium access control {(MAC)} and physical layer {(PHY)} specifications,''
  IEEE, 2020.

\bibitem{SepGozCol:J17}
M.~{Sepulcre}, J.~{Gozalvez}, and B.~{Coll-Perales}, ``Why 6 {Mbps} is not
  (always) the optimum data rate for beaconing in vehicular networks,''
  \emph{IEEE Transactions on Mobile Computing}, vol.~16, no.~12, pp.
  3568--3579, Dec 2017.

\bibitem{CamMolVinZha:J12}
C.~{Campolo}, A.~{Molinaro}, A.~{Vinel}, and Y.~{Zhang}, ``Modeling prioritized
  broadcasting in multichannel vehicular networks,'' \emph{IEEE Transactions on
  Vehicular Technology}, vol.~61, no.~2, pp. 687--701, Feb 2012.

\bibitem{BazZanMas:J19}
A.~Bazzi, A.~Zanella, and B.~M. Masini, ``Optimizing the resource allocation of
  periodic messages with different sizes in {LTE-V2V},'' \emph{IEEE Access},
  pp. 1--1, 2019.

\bibitem{BazCecZanMas:J18}
A.~Bazzi, G.~Cecchini, A.~Zanella, and B.~M. Masini, ``Study of the impact of
  {PHY} and {MAC} parameters in {3GPP C-V2V} mode 4,'' \emph{IEEE Access}, pp.
  1--1, 2018.

\bibitem{MolGozSep:C18}
R.~Molina-Masegosa, J.~Gozalvez, and M.~Sepulcre, ``Configuration of the
  {C}-{V2X} {M}ode 4 sidelink {PC}5 interface for vehicular communications,''
  in \emph{MSN 2018}.

\bibitem{TogSaiMahMugFalRaoDas:C18}
B.~Toghi, M.~Saifuddin, H.~N. Mahjoub, M.~Mughal, Y.~P. Fallah, J.~Rao, and
  S.~Das, ``Multiple access in cellular {V2X}: Performance analysis in highly
  congested vehicular networks,'' in \emph{2018 IEEE Vehicular Networking
  Conference (VNC)}.\hskip 1em plus 0.5em minus 0.4em\relax IEEE, 2018, pp.
  1--8.

\bibitem{ETSI_TS_103_613}
``Intelligent transport systems ({ITS}); access layer specification for
  intelligent transport systems using lte vehicle to everything communication
  in the 5,9 ghz frequency band,'' \emph{ETSI TS 103 613 V1.1.1}, 2018.

\bibitem{tong16}
Z.~Tong, H.~Lu, M.~Haenggi, and C.~Poellabauer, ``A stochastic geometry
  approach to the modeling of {DSRC} for vehicular safety communication,''
  \emph{IEEE Transactions on Intelligent Transportation Systems}, vol.~17,
  no.~5, pp. 1448--1458, 2016.

\bibitem{ZhaCheYanEtAl:J12}
W.~Zhang, Y.~Chen, Y.~Yang, X.~Wang, Y.~Zhang, X.~Hong, and G.~Mao, ``Multi-hop
  connectivity probability in infrastructure-based vehicular networks,''
  \emph{IEEE JSAC}, vol.~30, no.~4, pp. 740--747, May 2012.

\bibitem{BazZanCecMas:J19}
A.~{Bazzi}, A.~{Zanella}, G.~{Cecchini}, and B.~M. {Masini}, ``Analytical
  investigation of two benchmark resource allocation algorithms for
  {LTE-V2V},'' \emph{IEEE Transactions on Vehicular Technology}, vol.~68,
  no.~6, pp. 5904--5916, June 2019.

\bibitem{ParKimHon:J18}
Y.~Park, T.~Kim, and D.~Hong, ``Resource size control for reliability
  improvement in cellular-based {V2V} communication,'' \emph{IEEE Transactions
  on Vehicular Technology}, pp. 1--1, 2018.

\bibitem{3GPP_TR_36_885}
``Technical specification group radio access network; study on {LTE}-based
  {V2X} services,'' \emph{3GPP {TR} 36.885 V14.0.0}, July 2016.

\bibitem{3GPP_EN_302_637_2}
``Intelligent transport systems {(ITS)}; vehicular communications; basic set of
  applications; part 2: Specification of cooperative awareness basic service,''
  \emph{3GPP {EN} 302.637-2 V1.3.1}, September 2014.

\bibitem{Car2Car_CAMstats}
``Survey on {ITS-G5 CAM} statistics,'' CAR 2 CAR Communication Consortium,
  TR2052, 2018.

\bibitem{ETSI_TS_102_636}
``Intelligent transport systems ({ITS}); vehicular communications;
  {GeoNetworking}; part 4: Geographical addressing and forwarding for
  point-to-point and point-to-multipoint communications; sub-part 2:
  Media-dependent functionalities for {ITS-G5},'' \emph{ETSI TS 102 636-4-2},
  2013.

\bibitem{ETSI_EN_302_663}
``Intelligent transport systems {(ITS)}; {ITS-G5} access layer specification
  for intelligent transport systems operating in the 5 {GHz} frequency band,''
  \emph{ETSI EN 302 663}, 2020.

\bibitem{BarMasMarSarBaz:J21}
S.~{Bartoletti}, B.~M. {Masini}, V.~{Martinez}, I.~{Sarris}, and A.~{Bazzi},
  ``Impact of the generation interval on the performance of sidelink {C-V2X}
  autonomous mode,'' \emph{IEEE Access}, vol.~9, pp. 35\,121--35\,135, 2021.

\bibitem{ETSI_103_574}
``Intelligent transport systems {(ITS)}; congestion control mechanisms for the
  {C-V2X PC5} interface; access layer part,'' \emph{ETSI TS 103 574 V1.1.1},
  Nov. 2018.

\bibitem{TodBarCamMolBerBaz:21}
V.~Todisco, S.~Bartoletti, C.~Campolo, A.~Molinaro, A.~O. Berthet, and
  A.~Bazzi, ``Performance analysis of sidelink {5G-V2X} mode 2 through an
  open-source simulator,'' \emph{IEEE Access}, vol.~9, 2021.

\end{thebibliography}
